\documentclass[12pt,preprint]{aastex}
\usepackage{url}
\usepackage{lscape}
\usepackage{float}
\usepackage{amsmath}
\usepackage{natbib}
\usepackage{graphicx}
\usepackage[pdftex]{hyperref}
\begin{document}

\title{Sulfidic Anion Concentrations on Early Earth for Surficial Origins-of-Life Chemistry}

\author{Sukrit Ranjan\altaffilmark{1,2,*},  Zoe R. Todd\altaffilmark{1}, John D. Sutherland\altaffilmark{3},  Dimitar D. Sasselov\altaffilmark{1}}

\altaffiltext{1}{Harvard-Smithsonian Center for Astrophysics, Cambridge, MA 02138, USA}
\altaffiltext{2}{MIT Dept. of Earth, Atmospheric, and Planetary Sciences, Cambridge, MA 02139, USA}
\altaffiltext{3}{Medical Research Council Laboratory of Molecular Biology, Cambridge, CB2 0QH, UK}
\altaffiltext{*}{Corresponding author: 77 Massachusetts Avenue, Room 54-1719, Cambridge, MA 02139, USA; sukrit@mit.edu; t:617-253-6283, f:617-324-2055}

\date{\today}

\begin{abstract}
A key challenge in origin-of-life studies is understanding the environmental conditions on early Earth under which abiogenesis occurred. While some constraints do exist (e.g., zircon evidence for surface liquid water), relatively few constraints exist on the abundances of trace chemical species, which are relevant to assessing the plausibility and guiding the development of postulated prebiotic chemical pathways which depend on these species. In this work, we combine literature photochemistry models with simple equilibrium chemistry calculations to place constraints on the plausible range of concentrations of sulfidic anions (HS$^-$, HSO$_3^{-}$, SO$_3^{2-}$) available in surficial aquatic reservoirs on early Earth due to outgassing of SO$_2$ and H$_2$S and their dissolution into small shallow surface water reservoirs like lakes. We find that this mechanism could have supplied prebiotically relevant levels of SO$_2$-derived anions, but not H$_2$S-derived anions. Radiative transfer modelling suggests UV light would have remained abundant on the planet surface for all but the largest volcanic explosions. We apply our results to the case study of the proposed prebiotic reaction network of \citet{Patel2015}, and discuss the implications for improving its prebiotic plausibility.  In general, epochs of moderately high volcanism could have been especially conducive to cyanosulfidic prebiotic chemistry. Our work can be similarly applied to assess and improve the prebiotic plausibility of other postulated surficial prebiotic chemistries that are sensitive to sulfidic anions, and our methods adapted to study other atmospherically-derived trace species. %%%%190

\end{abstract}

%\maketitle
\maketitle

\section{Introduction}

A key challenge for origins-of-life studies is determining the environmental conditions on early Earth. Environmental conditions (e.g., pH, temperature, pressure, chemical feedstock abundance, etc.) play a major role in determining the kinds of prebiotic chemistry that are possible or probable, and hence can help constrain the plausibility of proposed origin-of-life scenarios (e.g, \citealt{Urey1952}, \citealt{Corliss1981}, \citealt{McCollom2013}, \citealt{Ruiz-Mirazo2014}). Consequently, it is critical to understand the range of environmental conditions available on the early Earth for abiogenesis to proceed. Work over the past few decades has begun to constrain the environmental conditions that may have been available for abiogenesis, including but not limited to the past presence of liquid water, the availability of UV light at the surface, the mix of gases being outgassed to the atmosphere, the bulk pH of the ocean, and the conditions available at deep-sea hydrothermal vents \citep{Bada1994, Farquhar2000, Mojzsis2001, Delano2001, Holm2001, McCollom2007, Trail2011, Mulkidjanian2012, Beckstead2016, Sojo2016, Ranjan2017a,Novoselov2017, Halevy2017}. 

One challenging environmental factor to constrain is the abundance of trace chemical species on early Earth. These species can be important to proposed prebiotic chemical pathways as feedstocks or catalysts, but their abundances on the early Earth can be difficult to determine due to their rarity and hence limited impact on an already scarce rock record. In this paper, we explore the plausible abundances of one such family of molecules: sulfidic anions, i.e. sulfur-bearing aqueous anions (e.g., hydrosulfide, HS$^-$; bisulfite, HSO$_3^-$; sulfite, SO$_3^{2-}$). Our initial interest in these molecules was stimulated by the role they play in the prebiotic chemistry proposed by \citet{Patel2015}, but our calculations are applicable to studies of surficial prebiotic chemistry in general. For discussion of the relevance of the surface environment and its attendant processes to prebiotic chemistry, see, e.g., \citet{Mulkidjanian2012, Walker2012, Mutschler2015, Forsythe2015, Rapf2016, He2017, Deamer2017}. Our results are not relevant to deep-sea origin-of-life scenarios, such as \citet{McCollom2007, LaRowe2008, Martin2008, Sojo2016}. 

We specifically explore the atmosphere as a planetary source for sulfidic anions through dissolution of volcanically outgassed SO$_2$ and H$_2$S in small, shallow aqueous reservoirs like lakes. The prebiotic Earth's atmosphere is thought to have been anoxic and more reducing than modern Earth \citep{Kasting2014}, and volcanism levels have been hypothesized to have been higher \citep{Richter1985}. Then the abundance of atmospheric H$_2$S and especially SO$_2$ should have been higher compared to modern day levels, and aqueous reservoirs in equilibrium with the atmosphere would have dissolved some of these gases in accordance with Henry's Law, forming sulfidic anions through subsequent dissociation reactions. We use simple equilibrium chemistry combined with literature photochemical modelling to estimate the concentrations of these sulfidic anions as a function of pSO$_2$ and pH$_2$S, and as a function of total sulfur outgassing flux. Elevated levels of atmospheric sulfur can lead to the formation of UV shielding gases and aerosols; consequently, we use radiative transfer calculations to constrain the surface UV radiation environment as a function of total sulfur outgassing flux. UV light is of interest to prebiotic chemists both as a potential stressor for abiogenesis \citep{Sagan1973, Cockell2000UVhist}, as a potential eustressor for abiogenesis \citep{Sagan1971, Mulkidjanian2003, Pascal2012, Sarker2013, Rapf2016, Xu2016}, and because of evidence that the nucleobases evolved in a UV-rich environment \citep{Rios2013, Beckstead2016}. 

We apply our calculations to the case study of the cyanosulfidic prebiotic systems chemistry of \citet{Patel2015}. Building on the work of \citet{Powner2009} and \citet{Ritson2012}, \citet{Patel2015} proposed a prebiotic reaction network for the synthesis of activated ribonucleotides, short sugars, amino acids and lipid precursors from a limited set of feedstock molecules in aqueous solution under UV irradiation (at 254~nm). This reaction network is of interest because of the progress it makes towards the longstanding problem of nucleotide synthesis, because it offers the promise of a common origin for many biomolecules, and because it imposes specific geochemical requirements on its environment, which can be compared against what was available on early Earth to constrain and improve the chemistry's prebiotic plausibility \citep{Higgs2015, Springsteen2015, Sponer2016}. Relevant to our work, the \citet{Patel2015} chemistry requires sulfidic anions to proceed, as both a photoreductant and as a feedstock for a subset of the network's reactions. \citet{Patel2015} proposed impactors as a source for the sulfidic anions; while possible, this scenario imposes an additional, local requirement for this chemistry to function. On the other hand, if the atmosphere could supply adequate sulfidic reductant (and feedstock) on a global basis, it would reduce the requirements for parts (or all) of this reaction network to function, and would make it more compelling as an origins-of-life scenario. We evaluate this scenario. While our paper focuses on the chemistry of \citet{Patel2015} as a case study, our work can be used to evaluate and improve the plausibility of any proposed sulfidic anion-sensitive surficial prebiotic chemistry. Our methods can be adapted to study the prebiotic surficial concentrations of other atmospherically-sourced aqueous species. 

\section{Background}
\subsection{Plausible Prebiotic Levels of H$_2$S and SO$_2$}
The abundances of H$_2$S and SO$_2$ in the Earth's atmosphere are set by photochemistry, and are sensitive to a variety of factors. One of the most important of these factors is the outgassing rate of these compounds from volcanoes into the atmosphere. Absent biogenic sources, atmospheric photochemistry models typically assume abiotic SO$_2$ outgassing rates of  $1-3\times10^{9}$ cm$^{-2}$~s$^{-1}$ \citep{Kasting1989, Zahnle2006, Hu2013, Claire2014}, consistent with the measured modern mean volcanogenic SO$_2$ outgassing rate of $1.7-2.4\times10^{9}$ cm$^{-2}$~s$^{-1}$ \citep{Halmer2002}. H$_2$S emission rates are indirectly estimated and much less certain; they range from $3.1\times10^{8}-7.7\times10^{9}$ cm$^{-2}$~s$^{-1}$. A common assumption in atmospheric modelling is that SO$_2$ and H$_2$S are outgassed in a 10:1 ratio (e.g., \citealt{Zahnle2006, Claire2014}). 

The early Earth is often hypothesized to have been characterized by higher levels of volcanic outgassing compared to the modern Earth due to presumed higher levels of internal heat and tectonic activity. Models often assume that Archaean SO$_2$ outgassing rates were $\sim3\times$ modern \citep{Richter1985, Kasting1989, Zahnle2006}. However, \citet{Halevy2014} point out that during the emplacement of major volcanogenic features such as the terrestrial basaltic plains, sulfur outgassing rates as high as $10^{10}-10^{11.5}$ cm$^{-2}$~s$^{-1}$ are possible, with the upper limit on outgassing rate coming from estimates of sulfur flux during emplacement of the Deccan Traps on Earth \citep{Self2006}.

No firm constraints exist for SO$_2$ and H$_2$S levels on the prebiotic Earth. \citet{Kasting1989} modeled a plausible prebiotic atmosphere of 2 bars CO$_2$, 0.8 bar N$_2$ atmosphere under $0.75\times$ present-day solar irradiation to account for the effects of the faint young Sun at 3.9 Ga. \citet{Kasting1989} assumed that sulfur was outgassed entirely as SO$_2$ at a total sulfur outgassing flux of $\phi_S=3\times10^{9}$ cm$^{-2}$~s$^{-1}$ into an atmosphere overlying an ocean saturated in SO$_2$; this last condition favors accumulation of SO$_2$ in the atmosphere. \citet{Claire2014} modeled an atmosphere of 0.99 bar N$_2$ and 0.01 bar CO$_2$, under irradiation by the 2.5 Ga Sun, with an SO$_2$:H$_2$S outgassing ratio of 10:1, for $\phi_S=1\times10^{8}-1\times10^{10}$ cm$^{-2}$~s$^{-1}$. \citet{Hu2013} modeled an atmosphere consisting of 0.9 bar CO$_2$ and 0.1 bar N$_2$ under irradiation by the modern Sun, with an SO$_2$:H$_2$S emission ratio of 2, for $\phi_S=3\times10^{9}-1\times10^{13}$ cm$^{-2}$~s$^{-1}$. The SO$_2$ and H$_2$S mixing ratios calculated by these models are shown in Table~\ref{tbl:compare_sulfur_models}; these mixing ratios may be trivially converted to partial pressures by multiplying against the bulk atmospheric pressure. Note that the \citet{Claire2014} and \citet{Kasting1989} values are surface mixing ratios, while the \citet{Hu2013} values are  column-integrated mixing ratios. Since H$_2$S and SO$_2$ abundances tend to decrease with altitude due to losses from photochemistry, column-integrated mixing ratios should be somewhat less than the surface mixing ratio. However, since density also decreases with altitude, mixing ratios at lower altitudes are more strongly weighted in the calculation of column-integrated mixing ratios, so the column-integrated mixing ratio tends to be close to the surface mixing ratio.  

\begin{table}[h]
\centering
\caption{Mixing ratios of H$_2$S and SO$_2$ for different early Earth models in the literature and different $\phi_S$.\label{tbl:compare_sulfur_models}} 
\begin{tabular}{p{8 cm}p{3 cm}p{3 cm}}
 \hline
 Model & $r_{H_{2}S}$ & $r_{SO_{2}}$ \\
 \hline
\citet{Kasting1989}$^{a}$, $\phi_S=3\times10^{9}$ cm$^{-2}$~s$^{-1}$& $2\times10^{-10}$&$2\times10^{-9}$ \\
\citet{Claire2014}$^{a}$, $\phi_S=3\times10^{9}$ cm$^{-2}$~s$^{-1}$& $1\times10^{-11}$ & $5\times10^{-11}$\\
\citet{Hu2013}$^{b}$, $\phi_S=3\times10^{9}$ cm$^{-2}$~s$^{-1}$& $4\times10^{-10}$ & $3\times10^{-10}$\\
\citet{Claire2014}$^{a}$, $\phi_S=1\times10^{10}$ cm$^{-2}$~s$^{-1}$& $3\times10^{-11}$ & $1\times10^{-10}$\\
\citet{Hu2013}$^{b}$, $\phi_S=1\times10^{10}$ cm$^{-2}$~s$^{-1}$& $1\times10^{-9}$ & $9\times10^{-10}$\\
 \hline
\end{tabular}
\\
a: Surface mixing ratio\\
b: Column-integrated mixing ratio\\
\end{table}

These models broadly agree that SO$_2$ and H$_2$S levels were low and increase with sulfur emission rate, but their estimates for $r_{SO_{2}}$  and $r_{H_{2}S}$  disagree with each other by up to a factor of 400. The \citet{Hu2013} estimates are typically higher than the other estimates considered. The variation in these abundances demonstrates the sensitivity of SO$_2$ and H$_2$S levels to atmospheric parameters such as composition and deposition velocities. Of these models, we find \citet{Hu2013} best matches the current fiducial understanding of conditions on early Earth: an atmosphere dominated by CO$_2$ and N$_2$, with volcanic outgassing of both SO$_2$ and H$_2$S, with oceans not saturated in SO$_2$ (as compared to possibilities for early Mars; see \citealt{Halevy2007}). \citet{Hu2013} also has the advantage of calculating atmospheric composition at higher values of sulfur outgassing flux than \citet{Kasting1989} and \citet{Claire2014}, encompassing the $1\times10^{11.5}$ cm$^{-2}$~s$^{-1}$ flux which is the upper limit of what \citet{Halevy2014} suggest possible for the emplacement of terrestrial basaltic plains. \citet{Hu2013} model processes including wet and dry deposition,  formation of H$_2$SO$_4$ and S$_8$ aerosol, and photochemistry and thermochemistry, with $>1000$ reactions included in their reaction network. We therefore use \citet{Hu2013} as a guide when estimating H$_2$S and SO$_2$ levels as a function of sulfur outgassing flux (see Appendix~\ref{sec:hutbl}), with the understanding that further, prebiotic-Earth specific modelling is required to constrain this relation with certainty. 

\section{Methods}
We consider a gas $Z$ dissolving into a surficial aqueous reservoir ($\lesssim$~1~m deep), through which the UV light required for prebiotic biomolecules synthesis can penetrate \citep{Ranjan2016}; our archetypal such environment is a shallow lake. To isolate the effects of atmospheric supply of $Z$, we assume no other source of $Z$ to be present (e.g., no geothermal source at the lake bottom). Henry's Law states that the concentration of $Z$, [Z], in aqueous solution at the air/water interface is proportional to the partial pressure of the gas at that interface. We assume the aqueous reservoir to be well-mixed and equilibrated throughout, so that the concentration of [Z] is uniform throughout the reservoir at the surficial value. If the reservoir is not well-mixed, then the dissolved gas concentration will vary deeper into the reservoir. Under our assumption of no non-atmospheric source of $Z$, [Z] would decrease with depth for a poorly-mixed aqueous reservoir. %The top layer of most bodies of water satisfies this well-mixed assumption (the so-called epilimnion, see \cite{Stepankp2016}). The well-mixed epilimnion is often meters thick in lakes on modern Earth, \cite{Keller2006}, and can encompass the entire lake, depending on parameters such as season \footnote{\url{http://www.watercenter.org/physical-water-quality-parameters/water-temperature/temperature-ranges-in-lakes/}, accessed 2015 May 23.}. 

This method of calculating [Z] is predicated on the assumption that the aqueous body is in equilibrium with the atmosphere, that is, that the solution is saturated in $Z$ and the sink and source of $Z$ is outgassing and deposition from the atmosphere. This assumption is valid when there are no other sinks to drive the system away from equilibrium. We discuss the veracity of this assumption in Section~\ref{sec:othersinks}. In brief, this assumption is valid for shallow, well-mixed lakes that are not very acidic or hot, but not valid for deep, acidic, or hot waters. For these scenarios, our calculations provide upper bounds on [Z].

In aqueous solution, H$_2$S undergoes the dissociation reactions

\begin{align}
H_2S &\rightarrow HS^- + H^+, pKa_{H_{2}S,1}=7.05 \\
HS^- &\rightarrow S^{2-}+H^{+}, pKa_{H_{2}S,2}=19
\end{align}

Where the pKa values are taken from \citet{CRC90}, and can be related to the corresponding equilibrium constants by $Ka_X=10^{-pKa_{X}}$. Similarly, SO$_2$ undergoes the reactions

\begin{align}
SO_2 + H_2O &\rightarrow HSO_3^- + H^+, pKa_{SO_{2},1}=1.86 \\
HSO_3^- &\rightarrow SO_3^{2-}+H^{+}, pKa_{SO_{2}, 2}=7.2\\
HSO_3^- + SO_2 &\rightarrow HS_2O_5^-, pKa_{SO_{2}, 3}=1.5
\end{align}

Where the $pKa$ values are from \citet{Neta1985}. 

To compute the abundances of these different sulfur-bearing compounds as a function of [Z], we must make assumptions as to the background chemistry of the aqueous reservoir they are dissolved in, especially its pH. If the reservoir is completely unbuffered (e.g., pure water), its pH (and hence the speciation of S-bearing compounds) will be completely determined by [Z]. At the other extreme, if the reservoir is completely buffered, its pH will be independent of [Z]. Natural waters typically lie in between these two extremes; they are often buffered by mineral or atmospheric interactions towards a certain pH\footnote{For example, the oceans on modern Earth are buffered to a pH of $8.1-8.2$ due primarily to carbonate buffering \citep{Hall-Spencer2008, Zeebe2009}; estimates of ancient ocean pH vary, but often invoke slightly lower pH due to posited higher CO$_2$ levels early in Earth's history (see, e.g., \citealt{Morse1998}, \citealt{Amend2009};  \citealt{Halevy2017} and sources therein). Smaller bodies, like lakes, can have an even wider range of pHs due to local conditions; lakes on modern Earth can have pH$<1$ (e.g., Kawah Ijen crater lake; \citealt{Lohr2005}), and pH$>11$ (e.g., Lake Natron; \citealt{Grant2000}).}, but with enough atmospheric supply their buffers can be overwhelmed. We explore these bracketing cases below, with the understanding that the true speciation behavior in nature was most likely somewhere in between.

\subsection{Calculating Dissolved Gas Concentration\label{sec:methods_henry}}
We use Henry's law, coupled with the well-mixed reservoir assumption, to calculate the concentration of molecules dissolved from the atmosphere. Henry's Law states that for a species $Z$, 

\begin{align}
[Z]=H_Z f_Z, 
\end{align}
where $H_Z$ is the gas-specific Henry's Law constant and $f_Z$ is the fugacity of the gas. Over the range of temperatures and pressures relevant to surficial prebiotic chemistry, the gases in our study are ideal, and consequently $f_Z=p_Z$, the partial pressure of $Z$. We make this simplifying assumption throughout our study. 

At $T_0=298.15$ K, the Henry's Law constants for H$_2$S and SO$_2$ dissolving in pure water are $H_{H_{2}S}=0.101$ M/bar and  $H_{SO_{2}}=1.34$ M/bar, respectively. Increasing salinity tends to decrease $H_G$, a process known as salting out. Similarly, increasing temperature also tends to decrease $H_C$. Our overall results are insensitive to variations in temperature of 25K from $T_0$ and $0\leq$[NaCl]$\leq1$ M; see Appendix~\ref{sec:appendix_henry} and Appendix~\ref{sec:appendix_henry_temp}. For simplicity, we therefore neglect the temperature- and salinity-dependence of Henry's Law.

\subsection{Unbuffered Solution}
Consider an unbuffered solution with dissolved $Z$, whose properties are determined entirely by the reactions $Z$ and its products undergo. From the definition of equilibrium constant, we can use the H$_2$S and SO$_2$ speciation reactions to write:

\begin{align}
\frac{a_{HS^{-}}a_{H^{+}}}{a_{H_{2}S}}=Ka_{H_{2}S,1} \label{eqn:h2s_1}\\
\frac{a_{S^{2-}}a_{H^{+}}}{a_{HS^{-}}}=Ka_{H_{2}S,2} \label{eqn:h2s_2}
\end{align}

and 

\begin{align}
\frac{a_{HSO3^{-}}a_{H^{+}}}{a_{SO_{2}}}=Ka_{SO_{2},1} \label{eqn:so2_1}\\
\frac{a_{SO3^{2-}}a_{H^{+}}}{a_{HSO3^{-}}}=Ka_{SO_{2},2} \label{eqn:so2_2}\\
\frac{a_{HS_{2}O_{5}^{-}}}{a_{SO_{2}}a_{HSO3^{-}}}=Ka_{SO_{2},3} \label{eqn:so2_3}
\end{align}

Where $a_C$ is the activity of species $C$. $a_C$ is related to the concentration of $C$, $[C]$, by $a_C=\gamma_C [C]$, where $\gamma_C$ is the activity coefficient \citep{Misra2012}. The use of activities instead of concentrations accounts for ion-ion and ion-H$_2$O interactions. $\gamma=1$ for a solution with an ionic strength of $I=0$. For ionic strengths of 0-0.1 M, we calculate the activity coefficients for each species as a function of solution ionic strength using Extended Debye-Huckel theory \citep{DebyeHuckel}. The activity coefficients in this formalism are calculated by:

\begin{equation}
log(\gamma_C) = -Az_C^2 \frac{I^{0.5}}{1+B\alpha_C I^{0.5}}
\end{equation}

Here, $A$ and $B$ are constants that depend on the temperature, density, and dielectric constant of the solvent; we use $A=0.5085$~M$^{-1/2}$ and $B=0.3281$~M$^{-1/2}$$\AA^{-1}$, corresponding to 25$^\circ$C water \citep{Misra2012} (our results are robust to this assumption; see Appendix~\ref{sec:appendix_temp}). $z_C$ is the charge of species $C$. $\alpha_C$ is an ion-specific parameter with values related to the hydration radius of the aqueous species; we took our $\alpha_C$ values from \citet{Misra2012}. We were unable to locate a value of $\alpha_C$ for HS$_{2}$O$_{5}^{-}$, and consequently take $\gamma_{HS_{2}O_{5}^{-}}=1$ throughout. $I$ is the ionic strength of the solution, defined as: %HS_{2}O_{5}^{-}

\begin{equation}
I=0.5 (\Sigma_C [C]z_C^2)
\end{equation}

We can combine these equations with the equation for water dissociation:
\begin{align}
H_2O \rightarrow OH^{-}+&H^{+},~pK_w=14\\
(a_{H^{+}})(a_{OH^{-}})&=K_w
\end{align}

and the requirement for charge conservation:

\begin{align}
\Sigma_C z_C [C] = 0
\end{align}

With [Z] specified by Henry's Law and our assumption of a well-mixed reservoir, this system is fully determined, and we can numerically solve it to determine the concentration of each of the species above as a function of pZ and $I$. A wide range of ionic strengths are possible for natural waters; modern freshwater systems like rivers have typical ionic strengths of order $1 \times 10^{-3} M$ \citep{PhysChemLakes}, whereas modern terrestrial oceans have an ionic strength of $0.7 M$~\footnote{\url{http://www.aqion.de/site/69}, accessed 29 November 2016}. The concentrations of divalent cations, especially Mg$^{2+}$ and Ca$^{2+}$, in early oceans has been suggested to be near 10mM \citep{DeamerDworkin}. A more fundamental constraint comes from vesicle formation, which is known to be inhibited at high salt concentrations and hence ionic strengths: \citet{Maurer2016} report that lipid vesicle formation is impeded in solutions with $I>0.1 M$. These considerations motivate our focus on low ionic strength waters, with $I\leq0.1 M$ \footnote{A further practical challenge with extending our calculations to higher ionic strengths is that the parameters required to compute the activity coefficients at high ionic strengths (e.g., via the \citealt{TJ1974} formalism) are not available for many of the species we consider.}. 

We calculate the speciation of sulfur-bearing species from dissolved H$_2$S and SO$_2$ for $I=0$ and $I=0.1 M$; the results are shown in Figs.~\ref{fig:h2s} and ~\ref{fig:so2}. $I=0$ is the lowest possible ionic strength, and $I=0.1M$ corresponds to the limit from lipid vesicle formation.

%%%%%%ZT updated gamma_X \geq here with new values. 
\subsection{Buffered Solution}
Consider now an aqueous reservoir that is buffered to a given pH. For example, the pH of the modern oceans is buffered by calcium carbonate to a global mean value of $8.1-8.2$ \citep{Hall-Spencer2008}. Then, we know [H$^{+}$], and can hence calculate the speciation of dissolved H$_2$S and SO$_2$ from the equilibrium constant equations \ref{eqn:h2s_1}-\ref{eqn:h2s_2} and \ref{eqn:so2_1}-\ref{eqn:so2_3} individually. Our results are insensitive to ionic strength for $I\leq0.1$M (see Figs. \ref{fig:h2s} and \ref{fig:so2}, and Appendix~\ref{sec:appendix_activitycoeffs}), and $I\leq0.1$M is required for vesicle formation and other prebiotic chemistry \citep{Maurer2016, Deamer2017}, motivating us to take $I=0$ for simplicity.  %\citet{Deamer2017}The geochemical scenario of \citet{Patel2015} invokes lakes and rivers, not oceans, motivating us to focus on low-ionic strength environments. For $I\leq1 \times 10^{-3}$, $\gamma_X\geq 0.868$ for the species we consider (\ref{sec:appendix_activitycoeffs}). We therefore take $I=0$ for simplicity for this calculation. 
%%need to confirm that ionic strength with updated activity coefficients doesn't affect results. 

With Henry's Law and our assumption of a well-mixed reservoir, we can readily calculate the concentration of the above species as a function of pH$_2$S or pSO$_2$ and pH. The results of this calculation are presented in Figs.~\ref{fig:h2s} and ~\ref{fig:so2} for three representative pHs. We selected pH=8.2, corresponding to modern ocean; pH=7, corresponding to the near-neutral phosphate-buffered conditions in which \citet{Patel2015} conducted their experiments; and pH=4.25, corresponding to raindrops in a pCO$_2\sim0.1$ bar atmosphere \citep{Halevy2007}. Such high CO$_2$ levels are hypothesized for the young Earth in order to power a greenhouse effect large enough to maintain clement surface conditions \citep{Kasting1993}.

The code used to implement these calculations is available for validation and extension at \url{https://github.com/sukritranjan/RanjanToddSutherlandSasselov2017.git}. 

%%%\subsection{Sensitivity of Results to Assumptions}
%%%%In the above calculations, we have implicitly assumed that the atmospheric SO$_2$ and H$_2$S concentrations are not significantly depleted by the dissolution and speciation of these gases in small, shallow aqueous reservoirs (like lakes) of the type we consider. This is equivalent to assuming that the atmospheric SO$_2$ and H$_2$S inventories are large compared to the inventories of SO$_2$ and H$_2$S drawn down by lakes. The atmospheric inventory of a gas $Z$ can be quantified as $(4\pi R_{\Earth}^2)(N_{Z})$, where $R_{\Earth}$ is the radius of the Earth, and $N_{Z}$ is the column density of $Z$ in the atmosphere; $N_{Z}$ can be quantified by $N_{Z}=r_Z\frac{P}{\mu g}$, where $P$ is the surface pressure, $\mu$ is the mean molecular mass of the atmosphere, and $g$ is the acceleration due to gravity. The lake inventory of $Z$ can be calculated by $(r_Z P)(H_Z) (A_{lake} h_{lake})$; the surface area of natural lakes on modern Earth is $A_{lake}=4.2\times10^{16}$ cm$^2$ \citep{Downing2006}. Only the top few meters of lakes are well-mixed \citep{Keller2006,Stepanenko2016}; for conservatism, we take $h_{lake}=1\times10^{3}$ cm. 
%%%
%%%Combining these expressions, we find that the ratio between the 

\section{Results}

\subsection{H$_2$S vs SO$_2$\label{sec:versus}}
\begin{figure}[h]
\centering
\includegraphics[width=0.8\linewidth]{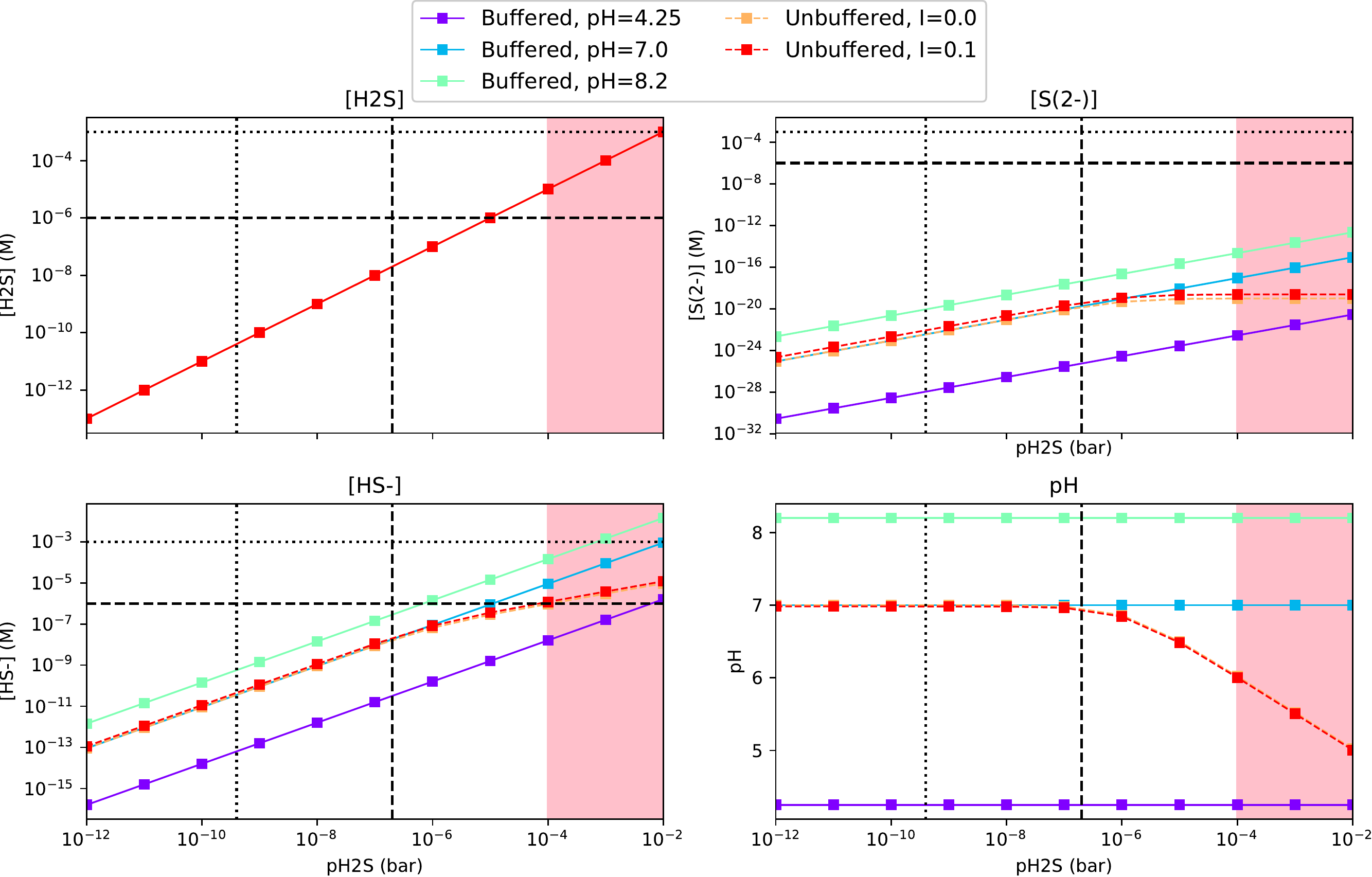}
\caption{Concentrations of sulfur bearing compounds and pH as a function of pH$_2$S for a well-mixed aqueous reservoir. [H$_2$S] is calculated from Henry's Law; the concentrations of HS$^-$ and S$^{2-}$ are calculated from equilbrium chemistry for 1) solutions buffered to various pHs, and 2) unbuffered solutions with varying ionic strengths. The vertical dotted line demarcates the expected pH$_2$S for an abiotic Earth with a weakly reducing CO$_2$-N$_2$ atmosphere with modern levels of sulfur outgassing, from \citet{Hu2013}. The vertical dashed line demarcates the expected pH$_2$S for the same model, but with outgassing levels of sulfur corresponding to the upper limit of the estimate for the emplacement of the terrestrial flood basalts. In the red shaded area, pH$_2$S is so high it blocks UV light from the planet surface, meaning UV-dependent prebiotic pathways, e.g., those of \citet{Patel2015}, cannot function \citep{Ranjan2017a}. The red curve largely overplots the orange, demonstrating the minimal impact of ionic strength on the calculation for $I\leq 0.1$. The horizontal dashed and dotted lines demarcate micromolar and millimolar concentrations, respectively. The cyanosulfidic chemistry of \citet{Patel2015} has been demonstrated at millimolar S-bearing photoreductant concentrations, and at least high micromolar levels of these compounds are thought to be required for high-yield prebiotic chemistry.\label{fig:h2s}} 
\end{figure}

Fig.~\ref{fig:h2s} shows the speciation of sulfur-bearing compounds from dissolved H$_2$S for an unbuffered reservoir, and reservoirs buffered to various pHs. Over the range of ionic strengths considered, HS$^-$ is the dominant anion, and S$^{2-}$ is present at negligible concentrations. As pH$_2$S increases, the pH of the unbuffered reservoir drops, but slowly. This is expected, since H$_2$S is a weak acid.

\begin{figure}[h]
\centering
\includegraphics[width=0.8\linewidth]{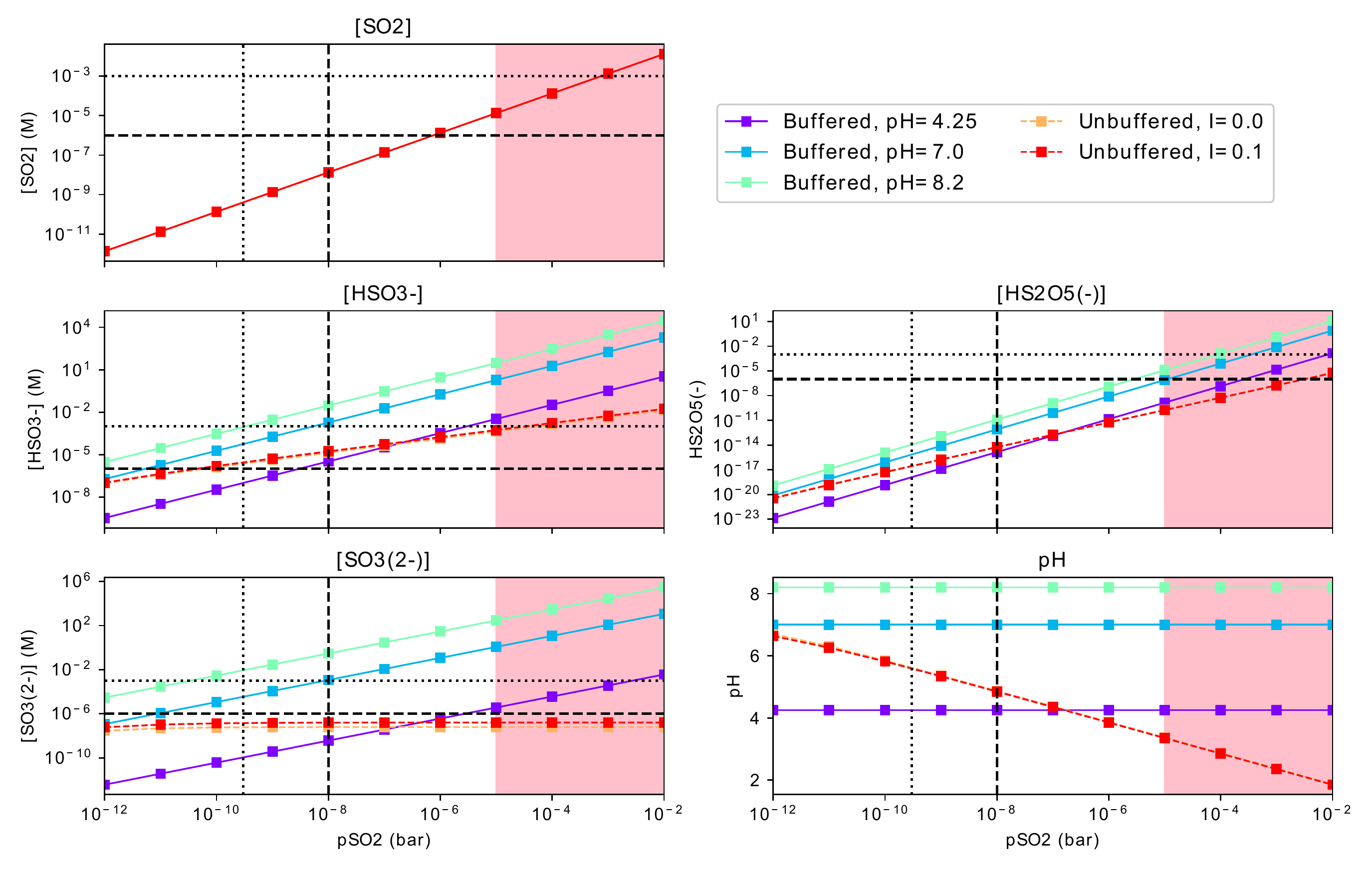}
\caption{Concentrations of sulfur bearing compounds and pH as a function of pSO$_2$ for a well-mixed aqueous reservoir. [SO$_2$] is calculated from Henry's Law; the concentrations of HSO$_3^-$, SO$_3^{2-}$, and HS$_2$O$_5^-$ are calculated from equilbrium chemistry for 1) solutions buffered to various pHs, and 2) unbuffered solutions with varying ionic strengths. The vertical dotted line demarcates the expected pSO$_2$ for an abiotic Earth with a weakly reducing CO$_2$-N$_2$ atmosphere with modern levels of sulfur outgassing, from \citet{Hu2013}. The vertical dashed line demarcates the expected pSO$_2$ for the same model, but with outgassing levels of sulfur corresponding to the upper limit of the estimate for the emplacement of the terrestrial flood basalts. In the red shaded area, pSO$_2$ is so high it blocks UV light from the planet surface, meaning UV-dependent prebiotic pathways, e.g., those of \citet{Patel2015}, cannot function \citep{Ranjan2017a}. The red curve largely overplots the orange, demonstrating the minimal impact of ionic strength on the calculation for $I\leq 0.1. $ The horizontal dashed and dotted lines demarcate micromolar and millimolar concentrations, respectively. The cyanosulfidic chemistry of \citet{Patel2015} has been demonstrated at millimolar S-bearing photoreductant concentrations, and at least high micromolar levels of these compounds are thought to be required for high-yield prebiotic chemistry. \label{fig:so2}}
\end{figure}

Fig.~\ref{fig:so2} shows the speciation of sulfur-bearing compounds from dissolved SO$_2$ for an unbuffered reservoir, and reservoirs buffered to various pHs. Because of the lack of O$_2$ in this anoxic era, the first dissociation of SO$_2$ forms sulfite, rather than sulfate. HSO$_3^-$ and SO$_3^{2-}$ are present at comparable levels; HS$_2$O$_5^-$ is negligible. As pSO$_2$ increases, the pH of the unbuffered reservoir falls off rapidly; this is expected since hydrated SO$_2$ is a strong acid. 

SO$_2$ is an order of magnitude more soluble than H$_2$S, and its first dissociation is much more strongly favored ($pKa_{SO_{2},1}=1.86$ vs $pKa_{H_{2}S,1}=7.05$). Consequently, far higher concentrations of sulfidic anions can be sustained for a given pSO$_2$ than for the same pH$_2$S (see Figs. \ref{fig:h2s} and \ref{fig:so2}). Maintaining micromolar concentrations of HS$^-$ requires pH$_2$S$\geq1\times10^{-6}$ bar at pH=8.2 (modern ocean), and pH$_2$S$\geq1\times10^{-5}$ bar for more neutral pHs. Maintaining micromolar concentrations of S$^{2-}$ is impossible over plausible ranges of pH and sulfur outgassing flux ($pKa_{H_{2}S,2}=19$). The concentration of sulfidic anions could be increased by going to higher pH and salinity. However, the reactions of, e.g., \citet{Patel2015} have not been demonstrated to proceed under such conditions.

By contrast, dissolved SO$_2$ gives rise to comparatively high concentrations of sulfidic anions due to higher solubility and a more favorable first ionization. Micromolar concentrations of HSO$_3^-$ are possible for pSO$_2>1\times10^{-11}$ bar for all but very acidic solutions; micromolar concentrations of SO$_3^{2-}$ are possible for solutions buffered to pH$\geq7$ over the same range. \emph{Millimolar} levels of HSO$_3^-$ and SO$_3^{2-}$ are possible for solutions buffered to pH$\geq8.2$ for pSO$_2\gtrsim10^{-10}$ bar, and for pH$\geq7$ solutions for pSO$_2\gtrsim10^{-8}$ bar. pSO$_2\geq3\times10^{-10}$ bar is expected for outgassing rates corresponding to the steady-state on early Earth according to the model of \citet{Hu2013} ($\phi_S=3\times10^{9}$ cm$^{-2}$~s$^{-1}$ ). During transient epochs of intense volcanism such as the emplacement of basaltic plains, emission rates might have risen as high as $\phi_S=10^{11.5}$ cm$^{-2}$~s$^{-1}$ \citep{Halevy2014, Self2006}, corresponding to pSO$_2=1\times10^{-8}$ bar. We note that estimates based on \citet{Hu2013} are for column-integrated abundances, and the surface abundances were likely modestly larger. Hence, it seems likely that the atmosphere could have supplied micromolar-levels of SO$_2$-derived anions for prebiotic chemistry, and perhaps even millimolar concentrations if the solution were buffered to slightly alkaline pH (e.g., pH comparable to the modern ocean).

\subsection{H$_2$S and SO$_2$\label{sec:and}}
\begin{figure}[h]
\centering
\includegraphics[width=.8\linewidth]{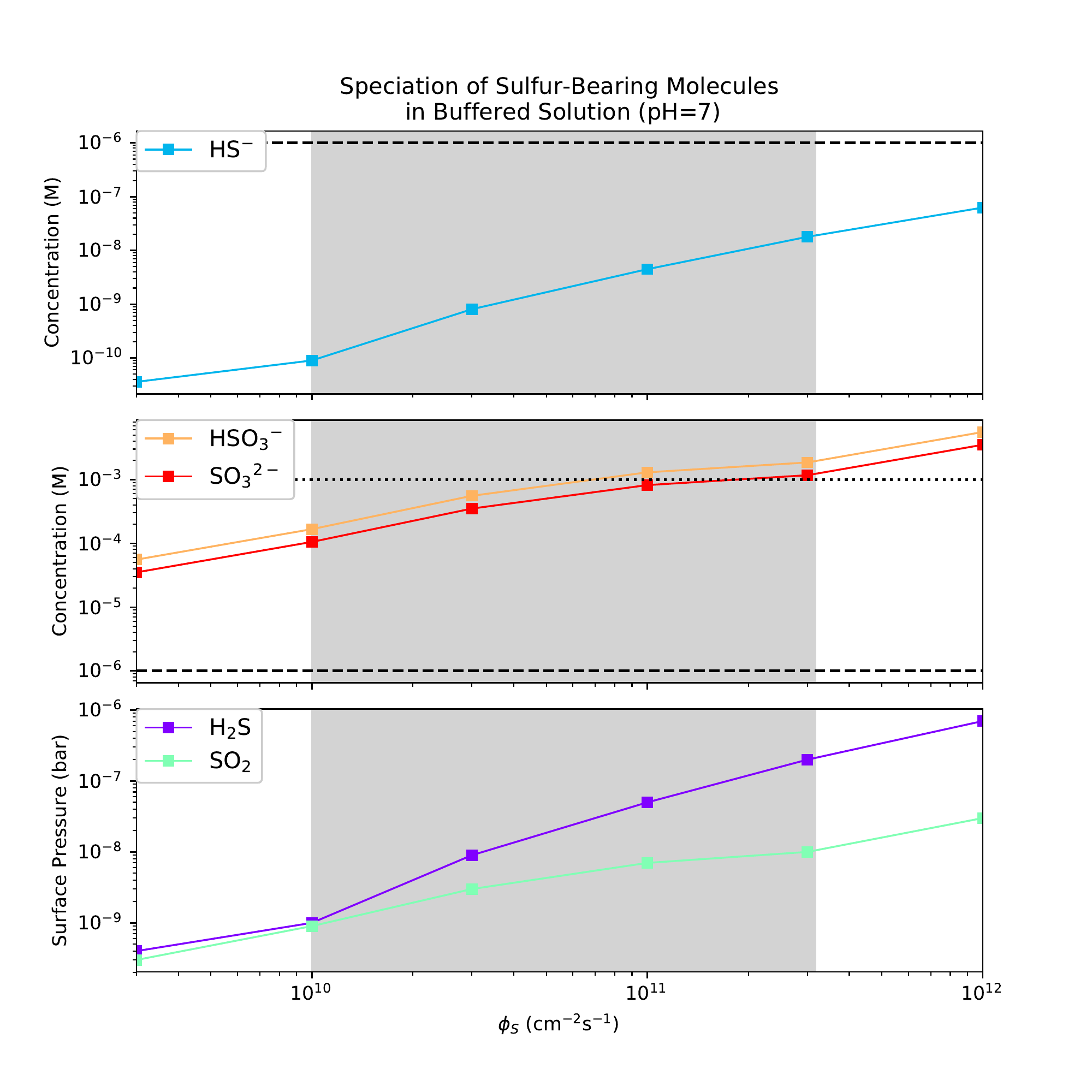}
\caption{Speciation of sulfur-bearing molecules in an aqueous reservoir buffered to pH=7 as a function of total sulfur emission flux $\phi_S$. The range of $\phi_S$ highlighted by \citet{Halevy2014} for emplacement of basaltic plains on Earth is shaded in grey. Horizontal dashed and dotted lines demarcate micromolar and millimolar concentrations, respectively. \label{fig:so2_h2s}}
\end{figure}

In Section~\ref{sec:versus} we evaluated the prospects for buildup of sulfur-bearing anions from dissolved atmospheric H$_2$S and SO$_2$ in isolation. However, H$_2$S and SO$_2$ are injected simultaneously into the atmosphere by volcanism, and would have been present at the same time. Fig.~\ref{fig:so2_h2s} presents the speciation of sulfur-bearing molecules from dissolved atmospheric H$_2$S and SO$_2$ in a solution buffered to pH$=7$ as a function of total sulfur outgassing rate, $\phi_S$. This pH corresponds approximately to the phosphate-buffered conditions in which the chemistry of \citet{Patel2015} proceeded\footnote{It is thought that these chemistries should proceed over a broad range of pH. However, they will proceed best for pH$\lesssim9.2$ (so that HCN tends to remain protonated) and pH$\gtrsim7$ (so that the sulfidic anions tend to remain deprotonated)}.  If the solution were buffered to higher pH sulfidic anion concentrations would be higher due to a more favorable first dissociation, and vice versa. 

As before, we connected the H$_2$S and SO$_2$ abundances connected to $\phi_S$ by the high-CO$_2$ model calculations of \citet{Hu2013}. We took the surface mixing ratio of these gases to equal the column-integrated mixing ratio, which may slightly underestimate the surface mixing ratio of these gases. $\phi_S=1-3\times10^{9}$ cm$^{-2}$~s$^{-1}$ for modern Earth, and  $\phi_S=10^{10}-10^{11.5}$ cm$^{-2}$~s$^{-1}$ have been suggested on a transient (1-10 year) basis for major volcanic episodes like the emplacement of basaltic plains on Earth \citep{Halevy2014, Self2006}. As discussed in Section~\ref{sec:versus}, SO$_2$-derived anions can build to micromolar levels at modern outgassing rates, and can build to millimolar levels during volcanic episodes like the emplacement of basaltic plains, while H$_2$S-derived anions cannot, absent highly alkaline conditions. %. In practice, the surface mixing ratio should be slightly higher than the column-integrated mixing ratio. Consequently, the pH$_2$S and pSO$_2$ we use here are lower bounds to the true values.

\subsection{Coupling to the UV Surface Environment\label{sec:uv}}
H$_2$S, SO$_2$, and their photochemical aerosol by-products (S$_8$, H$_2$SO$_4$) are robust UV shields, and at elevated levels their presence can dramatically reduce surface UV radiation \citep{Hu2013, Ranjan2017a}. This effect could be good for origin-of-life scenarios which do not require UV light, since  UV light can photolytically destroy newly formed biomolecules (e.g., \citealt{Sagan1973}). On the other hand, it could be bad for UV-dependent prebiotic chemistry, which depend on UV light to power their syntheses (e.g.,\citealt{Ritson2012}, \citealt{Patel2015}, \citealt{Xu2016}). In the latter case, it begs the question whether the elevated levels of SO$_2$ and H$_2$S that could supply the sulfidic anions required for cyanosulfidic chemistry might also quench the UV radiation also required by these pathways.

To explore this question, we calculated the attenuation of incoming 3.9 Ga solar radiation  (calculated from the models of \citealt{Claire2012}) by an CO$_2$-N$_2$-SO$_2$-H$_2$S atmosphere, using a two-stream radiative transfer model \citep{Ranjan2017a, Ranjan2017b}. We set the solar zenith angle to $48.2^\circ$, corresponding to the insolation-weighted mean value \citep{Cronin2014}, and the albedo to 0.2, a representative value for rocky planets consistent with past modelling\footnote{Our results are insensitive to the precise choice of albedo or solar zenith angle} \citep{Segura2003, Rugheimer2015}. We once again used the work of \citet{Hu2013} to connect H$_2$S and SO$_2$ abundances to $\phi_S$, and for consistency we assumed inventories of CO$_2$ and N$_2$ matching those assumed by \citet{Hu2013} (their high-CO$_2$ case). Our radiative transfer calculations are insensitive to the atmospheric T/P profile, because atmospheric emission is negligible at UV wavelengths and our UV cross-sections vary minimally as a function of temperature \citep{Ranjan2017b}; consequently, we assume a simple exponential profile to the vertical number density of the atmosphere. We also used the work of \citet{Hu2013} to estimate the total S$_8$ and H$_2$SO$_4$ aerosol loading in the atmosphere for each $\phi_S$, and calculated aerosol optical parameters using the same size distributions and complex indices of refraction as they did. Lacking detailed atmospheric profiles of the aerosol abundance as a function of altitude, we assumed the aerosols were distributed exponentially, with a scale height equal to the bulk atmosphere scale height (i.e. well-mixed). In practice, sulfur aerosols tend to form photolytically at higher altitudes, meaning our approach places more aerosol at low altitude and less aerosol at high altitude. Since the radiative impact of aerosol absorption is amplified lower in the atmosphere due to enhanced scattering, this means our treatment should slightly overestimate UV attenuation due to aerosols. Similarly, \citet{Hu2013} assume an aerosol size distribution with surface area mean diameter $D_S=0.1 \mu m$, at the lower end of the plausible $D_S=0.1-1 \mu m$ range, which maximizes the possible radiative impact of the sulfur aerosols. Consequently, our results should be interpreted as a lower bound on the true UV fluence. 

\begin{figure}[h]
\centering
\includegraphics[width=.8\linewidth]{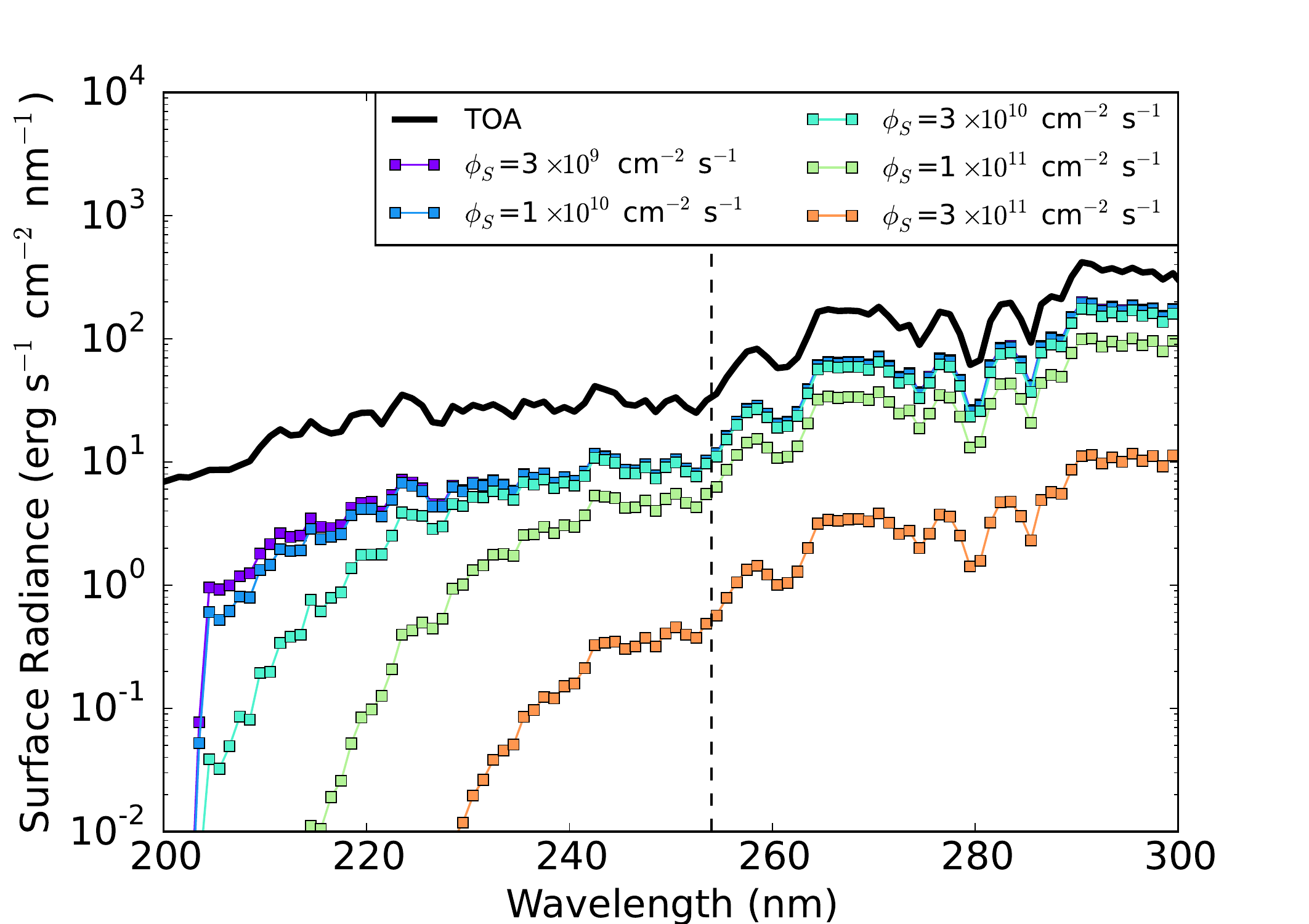}
\caption{UV surface radiance for the early Earth as a function of $\phi_S$, using the models of \citet{Hu2013}. The black solid line indicates the top-of-atmosphere (TOA) flux, i.e. the irradiation incident at the top of the atmosphere from the young Sun. The vertical dashed line demarcates 254 nm, the wavelength at which the low-pressure mercury lamps commonly used in prebiotic chemistry experiments emit. \label{fig:so2_h2s_trans}}
\end{figure}

Fig.~\ref{fig:so2_h2s_trans} presents the UV fluence available on the surface of the prebiotic Earth as a function of $\phi_S$ under these assumptions. For $\phi_S\leq 1\times10^{11}$ cm$^{-2}$~s$^{-1}$, UV radiation remains abundant on the planet surface. Millimolar levels of SO$_3^{2-}$ and HSO$_3^{-}$ are available in aqueous reservoirs buffered to $pH\geq 7$ for $\phi_S=1\times10^{11}$ cm$^{-2}$~s$^{-1}$. Consequently, volcanism could supply prebiotically relevant levels of SO$_3^{2-}$ and HSO$_3^{-}$ without blocking off the UV radiation required by UV-dependent prebiotic pathways for sulfur emission fluxes up to $\phi_S\leq1\times10^{11}$ cm$^{-2}$~s$^{-1}$ (near the upper edge of what is considered plausible for major terrestrial volcanic episodes). On the other hand, for $\phi_S\geq 3\times10^{11}$ cm$^{-2}$~s$^{-1}$, atmospheric sulfur-bearing gases and aerosols, especially the UV-absorbing S$_8$,  suppress surface UV radiation by an order of magnitude or more; this paucity of UV radiation may pose a challenge for UV-dependent prebiotic chemistry, but could create a very clement surface environment for UV-independent prebiotic chemistries. If one accepts the idea that the nucleobases show evidence of UV selection pressure \citep{Crespo-Hernandez2004, Serrano-Andres2009, Rios2013, Beckstead2016, Pollum2016}, this suggests the biogenic nucleobases evolved in an epoch with $\phi_S\leq 1\times10^{11}$ cm$^{-2}$~s$^{-1}$.

\section{Discussion}
\subsection{Sulfidic Anion Concentrations in Surficial Waters on Early Earth}
We have shown that terrestrial volcanism could have globally supplied the sulfidic anions SO$_3^{2-}$ and HSO$_3^{-}$, derived from the dissolution of SO$_2$ into aqueous solution, to shallow surficial aqueous reservoirs on early Earth. These compounds would have been available at micromolar levels for volcanic outgassing rates comparable to  the modern day. During episodes of high volcanism, such as those responsible for emplacement of basaltic plains ($\phi_S\approx1\times10^{11}$ cm$^{-2}$ s$^{-1}$), these compounds could have built up to the millimolar levels in shallow aqueous reservoirs buffered to pH$\geq7$. On the other hand, due to its lower solubility and unfavorable first dissociation, sulfidic anions derived from dissolving atmospheric H$_2$S can only be supplied at low concentrations (sub-micromolar) across the plausible range of pH$_2$S and pH\footnote{Our results are relevant to shallow, aqueous bodies of water, like lakes. By contrast, in the ocean volcanoes vent directly into the water, and the turnover time can be long. Consequently, one could envision significant buildup of H$_2$S near deep-ocean volcanoes. This scenario is beyond the scope of this work, but may be worthy of consideration for HS$^-$-dependent chemistry}. Therefore, other mechanisms must be invoked for supply of such anions, if required by a proposed prebiotic chemical pathway.

We conducted our calculations assuming a temperature of $T=25^\circ$C. We investigate the sensitivity of our results to temperatures ranging from $T=0-50^\circ$C in Appendix~\ref{sec:appendix_temp}, including temperature effects on both the reaction rate and the Henry's Law coefficient. While H$_2$S-derived anion concentrations are not significantly affected by temperature variations in this range, SO$_2$-derived anion concentrations are. This is because $H_{SO_{2}}$  decreases with temperature and pKa$_{SO_{2},1}$ increases with temperature\footnote{The exothermic first dissociation of SO$_2$ is disfavored at higher temperatures.}; both effects favor increased concentrations of HSO$_3^-$ and its derivatives, assuming a not-highly acidic ($pH>2.5$) solution. We find that while our overall conclusions are unchanged, concentrations of the SO$_2$-derived anions HSO$_3^-$ and SO$_3^{2-}$ are an order of magnitude higher for $T\approx0^\circ$C relative to $T=25^\circ$C, and an order of magnitude lower for $T\approx50^\circ$C, assuming a near-neutral reservoir. Consequently, cooler waters are more favorable environments for prebiotic chemistry which invokes HSO$_3^-$ or SO$_3^{2-}$. 

Sulfur-bearing gases and aerosols, in particular S$_8$, are strong UV absorbers, and if present at high enough levels could suppress UV-sensitive prebiotic chemistry. For $\phi_S\leq1\times10^{11}$ cm$^{-2}$ s$^{-1}$, corresponding to most of the plausible range of sulfur emission fluxes on early Earth, surface UV fluxes (200-300 nm) are not significantly attenuated by atmospheric absorbers, meaning that in the steady state and for most volcanic eruptions, abundant UV light should have reached the Earth's surface to power UV-dependent prebiotic chemistry. However, for the very largest volcanic eruptions, corresponding to the uppermost end of the plausible range of sulfur outgassing fluxes during terrestrial basaltic flood plain emplacement ($\phi_S=3\times10^{11}$ cm$^{-2}$ s$^{-1}$), surface UV fluence (200-300 nm) may be reduced by an order of magnitude or more. Hence, the very largest volcanic events\footnote{However, even in this case there may be a window in which prebiotic chemistry experiences both elevated SO$_2$ levels and plenty of UV light, since aerosol formation is not instantaneous. Detailed photochemical modelling is required to determine the timescale of aerosol formation after a large volcanic eruption on early Earth; absent such modelling, we note that on modern Earth, formation of sulfate aerosols from volcanic eruptions occurs on a timescale of weeks \citep{Robock2000}; we may speculate a similar timescale for aerosol formation on early Earth.} might create an especially clement surficial environment for UV-independent prebiotic chemistry% 

These results were derived using the high-CO$_2$ model of \citet{Hu2013}, which, while plausible, assumes more CO$_2$ and less N$_2$ than other models of prebiotic Earth (e.g., \citealt{Rugheimer2015}), and is hence comparatively oxidizing. We explored the sensitivity of our results to this assumption via the the N$_2$-rich model of \citet{Hu2013}. This model assumes 1 bar of N$_2$ and negligible CO$_2$, and is hence an unrealistic approximation to the early Earth, because an appreciable CO$_2$ inventory is expected due to climate constraints \citep{Kasting1993, Wordsworth2013h2}, and due to volcanic outgassing of CO$_2$. Hence, this model serves as an extreme bounding case. Assuming this model, we find that H$_2$S and SO$_2$ levels are lower than for the high-CO$_2$ case. SO$_2$-derived anions remain available at micromolar levels over the plausible range of $\phi_S$, but in order to build to millimolar levels require the assumption of reservoirs buffered to slightly alkaline pH (e.g., pH$\sim8.2$, modern ocean). HS$^-$ levels are even lower than in the CO$_2$-rich case. UV fluences are lower than in the CO$_2$-rich case, due to elevated levels of S$_8$ formation in this more reducing atmosphere; surface UV fluence (200-300 nm) is suppressed by an order of magnitude or more for $\phi_S\geq1\times10^{11}$ cm$^{-2}$ s$^{-1}$. Overall, this boundary case suggests that our finding that the atmosphere can supply prebiotically-relevant levels of  SO$_2$-derived anions but not H$_2$S-derived anions in conjunction with UV light remains true across a broad range of CO$_2$ and N$_2$ abundances, though both sulfidic anion abundances and UV are lower for more reducing, N$_2$-rich atmospheres. However a detailed exploration of the pCO$_2$-pN$_2$ parameter space with photochemical models is required to be certain of these findings.

%In this paper, we have focused on the concentrations of atmospherically-supplied SO$_2$ and H$_2$S derivatives in surficial aqueous solution. We note that our method should be adaptable for explorations of the abundances of other atmospherically-derived compounds, for studies of possible prebiotic environments and/or paleo-aqueous chemistry. 

\subsection{Impact of Other Sinks\label{sec:othersinks}}
Our analysis is predicated on the assumption that [Z] is set by Henry equilibrium, i.e. that the aqueous reservoir is saturated in H$_2$S and SO$_2$. This assumes no major sinks other than outgassing to the atmosphere. In this section, we examine the sensitivity of our results to this assumption. Microbial sinks (e.g., \citealt{Halevy2013}) are not relevant since we are concerned with prebiotic Earth; neither are oxic sinks, since the surface of early Earth was anoxic \citep{Kasting1981, Kasting1987, Farquhar2001, Pavlov2002, Li2013}. However, reactions with metal cations to produce insoluble precipitates and redox reactions could have been relevant; we explore these sinks. 

\subsubsection{Precipitation Reactions with Metal Cations\label{sec:cations}}
We explored the possibility that reactions of S-anions with metal cations might lead to formation of insoluble precipitates, which would act as a sink on S-anion concentrations. Such cations might have been delivered to aqueous reservoirs via weathering of rocks and minerals.

Under standard conditions, Fe$^{2+}$ and Cu$^{2+}$ react with H$_2$S(aq) to generate insoluble precipitates, like CuS and FeS$_2$ \citep{CRC98, Rickard2007}. Interaction of copper sulfides with cyanide solution can liberate HS$^-$ \citep{Coderre1999}, as invoked by \citet{Patel2015}. In general high-Cu/Fe waters (e.g. due to interaction with ores) will be even more HS$^-$-poor than we have modeled, with the caveat that specific local environmental factors (like the presence of aqueous cyanide) can prevent sulfide depletion due to precipitation. This reinforces our conclusion that HS$^-$ concentrations are unlikely to have reached prebiotically relevant levels on early Earth, absent unique local factors. For example, the aqueous cyanide required as a feedstock in the pathways of \citet{Patel2015} would also permit elevated HS$^{-}$ levels.

Ca$^{2+}$, produced by mineral weathering, reacts with sulfite to produce insoluble CaSO$_3$. Studying the Ca$^{2+}$-SO$_3^{2-}$ system requires considering the effects of carbonate (CO$_3^{2-}$) as well, because Ca$^{2+}$ forms precipitate with this anion as well, and because high levels of carbonate are expected in natural waters on early Earth due to elevated levels of atmospheric CO$_2$ required to solve the faint young Sun paradox \citep{Kasting1987}. While precisely modeling this geochemical system requires use of a geochemical model capable of accounting for all reactions involving sulfites and carbonates and their kinetics, we can get a first-order estimate of the impact of Ca$^{2+}$, as follows. Assuming parameters from \citet{Hu2013}, the flux of carbonates into solution due to deposition and speciation of atmospheric CO$_2$ is $r_{CO_{2}} n_{atm} v_{dep, CO_{2}}=2\times10^{15}$ cm$^{-2}$ s$^{-1}$ on the CO$_2$-rich early Earth, which exceeds the mean flux of Ca due to mineral weathering ($1-5\times10^{10}$ cm$^{-2}$ s$^{-1}$; \citealt{Taylor2012, Watmough2008}) by 5 orders of magnitude; thus, it is reasonable to assume the solution is saturated in CO$_2$ with abundance dictated by Henry's law of $(3.3\times10^{-2}$ M/bar)(0.9 bar) $= 0.03$M \citep{Sander2015}. Then, [CO$_3^{2-}$] =$(0.03$M$)(10^{7-6.35})(10^{7-10.33})=6\times10^{-5}$M at neutral pH (dissociation constants $Ka_{CO_{2},1}=6.35$ and $Ka_{CO_{2},1}=10.33$ from \citet{CRC98}\footnote{Using pKas for CO$_2$ dissociation from modern seawater, e.g., \citet{Zeebe2009}, results in much higher carbonate levels, much lower Ca levels, and a much higher threshold for CaSO$_3$ saturation, so this treatment is conservative.}). Since CaCO$_3$ ($K_{sp}=3.36 \times10^{-9}$ M$^{2}$, \citealt{CRC98}) is two orders of magnitude less soluble than CaSO$_3$ ($K_{sp}=3.1\times10^{-7}$ M$^{2}$, \citealt{CRC98}) and the sulfite flux is much less than the carbonate flux, we can assume that [Ca$^{2+}$] is dictated to first order by equilibrium with carbonate mineral, i.e.  [Ca$^{2+}$] $= 3.36\times10^{-9} $M$^2 /6\times10^{-5}$M$ = 6\times10^{-5}$ M. At this [Ca$^{2+}$], CaSO$_3^{2-}$ (s) will begin to form at [SO$_3^{2-}$]$= 3.1\times10^{-7}$ M$^{2}/6\times10^{-5}$M$ = 5\times10^{-3}$ M. The [SO$_3^{2-}$] we calculate does not exceed this threshold value across the plausible range of sulfur outgassing fluxes in our calculation, meaning the solution is unsaturated in CaSO$_3$ and precipitate does not form. Were pCO$_2$ lower, e.g.,  pCO$_2=0.2$ bar\footnote{Corresponding to the lower limit suggested by \citet{Kasting1987} from climate considerations.}, CaSO$_3$ precipitate formation begins at [SO$_3^{2-}$]$=1\times10^{-3}$M. However, if pH were low, the carbonate solubility would exceed sulfite solubility, and sulfite precipitates would form \citep{Halevy2007}; hence at low pH, sulfite and bisulfite concentrations will be below the values we calculate. Overall, our results are unaffected by CaSO$_3$ precipitation across most of parameter space, but CaSO$_3$ precipitation might be a significant sink on aqueous sulfite levels for acid solutions and/or for very low atmospheric CO$_2$-levels; calculations with a more thorough geochemical model (e.g., PHREEQC, \citealt{PHREEQC}) are required to constrain S-anion concentrations in this regime.

\subsubsection{Redox Reactions\label{sec:redox}}
We explored the possibility that redox reactions (disproportionation, comproportionation) might have acted as sinks to S-anion concentrations in shallow aqueous reservoirs on prebiotic Earth, or might otherwise affect the distribution of sulfidic anions. We identified the following reactions that are spontaneous near standard conditions \citep{Siu1999, Halevy2013}:

\begin{align}
4SO_3^{2-} +H^+\rightarrow 2SO_4^{2-} + S_2O_3^{2-} + H_2O \label{eqn:halevy} \\ 
2HS^- + 4HSO3^- \rightarrow 3S_2O_3^{2-}+3H_2O \label{eqn:siu}
\end{align} 

The kinetics of Reaction~\ref{eqn:halevy} are not well characterized near standard temperature, and are an active topic of research \citep{Mirzoyan2014, Amshoff2016}. \citet{Meyer1982} report sulfite and bisulfite are stable on timescales $\geq1$ year in anoxic conditions, while \citet{Guekezian1997} report decay of sulfite in days at pH$\geq12.8$. \citet{Halevy2013} propose that rate coefficients in the range $k_{\ref{eqn:halevy}}=\exp(\frac{-50 kJ mol^{-1}}{RT}) - \exp(\frac{-40 kJ mol^{-1}}{RT})$ s$^{-1}$ are plausible; at 293K, this corresponds to $1\times10^{-9} - 7\times10^{-8}$ s$^{-1}$, which correspond to timescales of $0.5 - 30$ years. The kinetics of Reaction~\ref{eqn:siu} have been determined as a function of temperature at pH=9 and $I=0.2$M by \citet{Siu1999}. At 293K, the rate coefficient is $k_{\ref{eqn:siu}}=4\times10^{3}$ M$^{-2}$ s$^{-1}$. At the S-anion concentrations relevant to our work\footnote{[HS$^-$]$\lesssim10^{-8}$M, [HSO$_3^-$]$\lesssim10^{-3}$M}, the timescale of this reaction is $\gtrsim1$ year. For comparison, putative prebiotic chemistry in laboratory studies often occurs on timescales of hours to days \citep[e.g., ]{Patel2015, Xu2017}.  %Moreover, at non-acidic pH, the resultant thiosulfate does not go to insoluble elemental sulfur, which is insoluble \citep{Johnston1973, Mizoguchi1976, Meyer1982};

We test the effects of redox reactions on S-anion concentrations by carrying out a dynamical equilibrium calculation for a shallow lake buffered to pH=7, with source the atmosphere and sink these redox reactions.  Following the treatment of \citet{Halevy2013}, the equilibrium equations can be written:
\begin{align}
r_{H_{2}S} n_{atm} v_{dep, H_{2}S} A_{catch} =(\frac{2}{3}k_{\ref{eqn:siu}}[HS^-][HSO_3^-]^2) A_{lake} d_{lake} \label{eqn:dynamic_h2s}\\
r_{SO_{2}} n_{atm} v_{dep, SO_{2}} A_{catch} =( \frac{4}{3}k_{\ref{eqn:siu}}[HS^-][HSO_3^-]^2 +  k_{\ref{eqn:halevy}}[S(IV)]) A_{lake} d_{lake} \label{eqn:dynamic_so2}
\end{align}

\begin{figure}[h]
\centering
\includegraphics[width=0.8\linewidth]{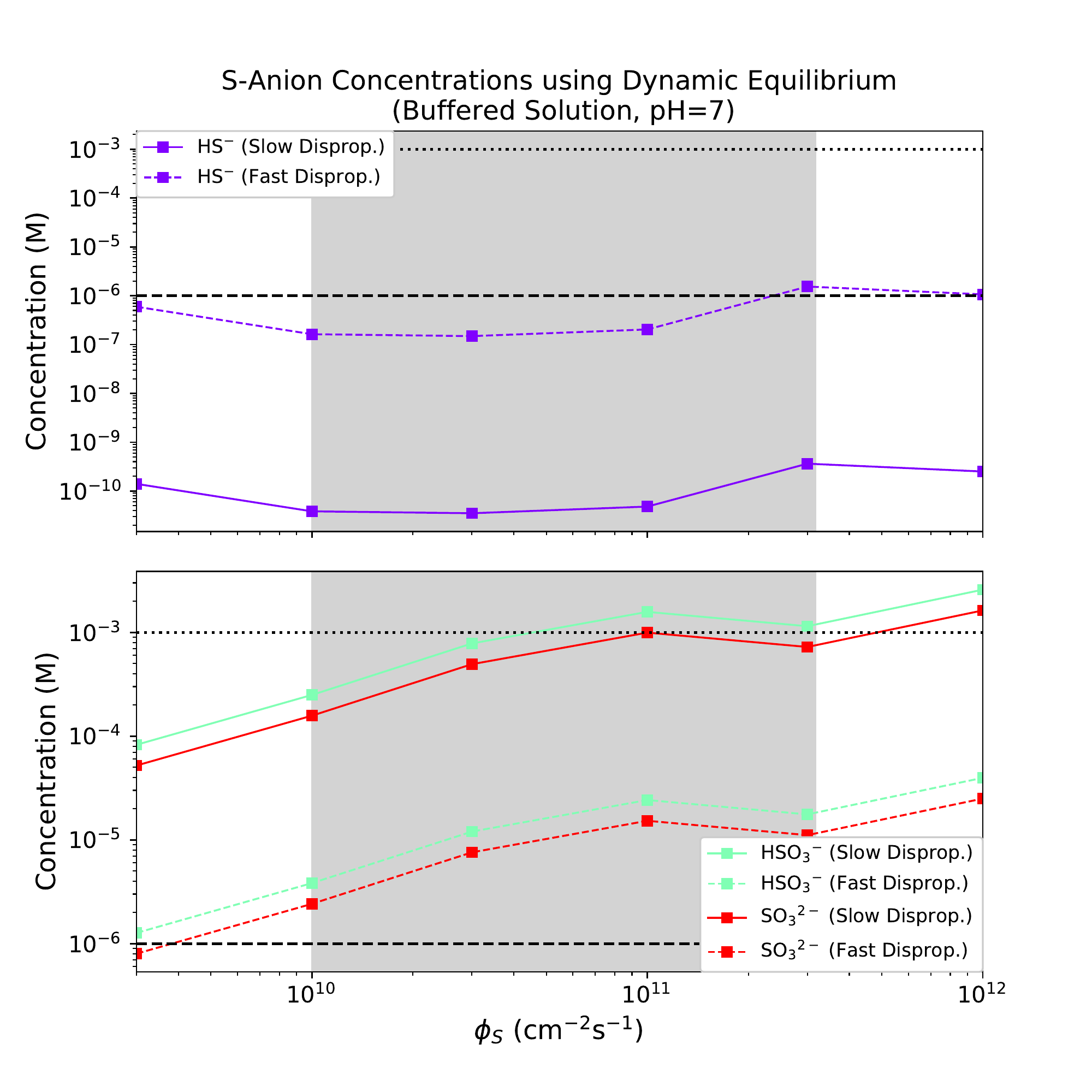}
\caption{Speciation of sulfur-bearing molecules in a shallow lake buffered to pH=7 as a function of total sulfur emission flux $\phi_S$, using a dynamic calculation with source atmospheric deposition and sink redox reactions. The range of $\phi_S$ highlighted by \citet{Halevy2014} for emplacement of basaltic plains on Earth is shaded in grey. Horizontal dashed and dotted lines demarcate micromolar and millimolar concentrations, respectively. [HS$^-$] would not be able to achieve the high concentrations calculated here for the slow disproportionation (low $k_{\ref{eqn:halevy}}$) case due to solubility constraints. \label{fig:redox}}
\end{figure}

For consistency with \citet{Hu2013}, we adopt $v_{dep, H_{2}S}=0.015$ cm s$^{-1}$, $v_{dep, SO_{2}}=1$ cm s$^{-1}$, $T=288$K, and $n_{atm}=\frac{1 bar}{kT}=2.4\times10^{19}$ cm$^{-3}$ . Since we are concerned with shallow, well-mixed lakes, we take the lake depth $d_{lake}=10^2$ cm. $A_{catch}$ is the catchment area of the lake and $A_{lake}$ is the surface area of the lake; we conservatively adopt $A_{catch}=A_{lake}$, which likely underestimates sulfur supply since the catchment area is often larger than the lake area. $[S(IV)]$ refers to the total concentration of S(IV) atoms in solution, and is calculated as $[S(IV)] = [SO_2] + [HSO_3^-] + [SO_3^{2-}] + 2[HS_2O_5^{-}]\approx [SO_2(aq)] + [HSO_3^-] + [SO_3^{2-}]$ \footnote{[HS$_2$O$_5^{-}$] is negligible for dilute [SO$_2$]}. Since we have specified  pH=7 and know the relevant pKas, we can calculate [HSO$_3^-$] from [S(IV)] and vice versa. With $r_{H_{2}S}$ and $r_{SO_{2}}$ specified from \citet{Hu2013}, we have a system of two equations in two variables that we can solve. Figure~\ref{fig:redox} shows the resultant S-anion concentrations as a function of $\phi_S$.

The dynamic calculation is very sensitive to the uncertainty in $k_{\ref{eqn:halevy}}$, with sulfite and bisulfite concentrations varying by 2 orders of magnitude and hydrosulfide concentrations varying by 4 across the range of $k_{\ref{eqn:halevy}}$ suggested by \citet{Halevy2013}. However, even with this uncertainty it is clear that prebiotically relevant levels ($\geq 1 \mu$M)of SO$_2$-derived anions are available across the range of plausible sulfur outgassing fluxes, with concentrations $\sim1-10\mu$M if sulfite disproportionation is fast and $\sim100-1000\mu$M if sulfite disproportionation is slow. Note depending on $k_{\ref{eqn:halevy}}$, it is possible for [HS$^-$] in the dynamic calculation to exceed the value calculated from solubility constraints; in reality, in well-mixed solution H$_2$S would de-gas when it reached the solubility limit, voiding equation~\ref{eqn:dynamic_h2s}. In these cases, [HS$^-$] is lower than the value calculated from the dynamic method, modestly increasing sulfite and bisulfite concentrations since Reaction~\ref{eqn:siu} is slower. S-anion concentrations increase as $d_{lake}$ and $T$ decrease, and are ultimately limited by gas solubility. Overall, our finding that prebiotically relevant levels of SO$_2$-derived anions were available in shallow well-mixed lakes on early Earth is robust to the effect of redox reactions, but it is possible for the precise concentrations to be lower than from our equilibrium calculation depending on the depth and temperature of the lake, and especially on the rate of sulfite disproportionation $k_{\ref{eqn:halevy}}$. Constraining $k_{\ref{eqn:halevy}}$ is key to improved modelling of abiotic sulfur chemistry.

\subsection{Case Study: Implications for Cyanosulfidic Systems Chemistry of \citet{Patel2015}}
The cyanosulfidic prebiotic chemistry of \citet{Patel2015} requires cyanide and sulfur-bearing anions, both as feedstocks and as sources of hydrated electrons through UV-driven photoionization. \citet{Patel2015} used HS$^-$ as their sulfidic anion, and propose impact-derived sources of metal sulfides (both from the impactor and from subsequent metallogenesis) and evaporatively concentrated iron sulfides as a source for HS$^-$. This postulated mechanism requires specific, local environmental conditions to function. By contrast, simple exposure of a non-acidic lake to the atmosphere anywhere on the planet would supply HSO$_3^-$ and SO$_3^{2-}$ at prebiotically relevant levels to either supplement the photochemical reducing capacity of HS$^-$, or function as sole sources of hydrated electrons in the \citet{Patel2015} chemistry. Indeed, recent work by the same group suggests that HSO$_3^-$  and SO$_3^{2-}$  can replace HS$^{-}$ as the source of hydrated electrons upon UV irradiation, and thus drive those parts of the reaction network that do not rely on HS$^{-}$ as a feedstocks (\citealt{Xu2017}, \emph{in prep}). Reducing or eliminating the dependence of the \citet{Patel2015} chemistry on HS$^-$ in favor of HSO$_3^-$ or SO$_3^{2-}$ increases the robustness of this chemistry, because no special local circumstances need to be invoked. This illustrates how geochemistry can inform improvements of the plausibility of prebiotic pathways.

Indeed, volcanism can be a source of more than sulfidic anions. Volcanism can also be a source of phosphates through partial hydrolysis of volcanically outgassed polyphospates \citep{Yamagata1991}, and a supplementary source of HCN through photochemical reprocessing of volcanically outgassed reducing species like CH$_4$ \citep{Zahnle1986}\footnote{Though some concentration mechanism would be required to achieve prebiotically relevant levels of HCN via this pathway}. Volcanism could thereby supply or supplement many of the C, H, O, N P \& S-containing feedstock molecules and photoreductants required by the \citet{Patel2015} chemistry. The UV light also required by the \citet{Patel2015} chemistry would be available at Earth's surface for all but the largest volcanic episodes ($\phi_S\geq3\times10^{11}$ cm$^{-2}$ s$^{-1}$). Hence, epochs of moderately high volcanism may have been uniquely conducive to cyanosulfidic prebiotic chemistry like that of \citet{Patel2015}, especially if they can be adapted to work with HSO$_3^-$ or SO$_3^{2-}$ instead of HS$^-$.

We considered alternate planetary sources for HS$^-$ for the \citet{Patel2015} chemistry. We explored whether shallow hydrothermal systems, such as hot springs, might provide prebiotically-relevant levels of HS$^-$. These sources are high-sulfur systems on modern Earth, and, if shallow, prebiotic chemistry in them might retain access to UV light while accessing high concentrations of sulfidic anions. Surveys of modern hydrothermal systems reveal examples of surficial systems that exhibit micromolar or even millimolar concentrations of HS$^{-}$ \citep{Xu1998, Vick2010, Kaasalainen2011}. However, high concentrations of HS$^{-}$ appear to only be achieved in hot systems\footnote{We speculate that HS$^{-}$-rich shallow hydrothermal systems tend to be hot because the same volcanism that supplies elevated levels of HS$^{-}$ also supplies elevated levels of heat.} ($T>60^\circ$ C, and typically higher). Similarly, studies of geothermal waters in Yellowstone National Park suggest sulfite availability at the $0.4-5 \mu$M level. However, such levels of sulfite were again accessed only in hot waters \citep{Kamyshny2014}. It is not clear how compatible such conditions are with prebiotic chemistry; for example, most of the cyanosulfidic chemistries of \citet{Patel2015} and \citet{Xu2016} were conducted at room temperature (25$^\circ$ C), and in general many molecules thought to be relevant to the origin of life, such as ribozymes, RNA and their components, are more stable and function better at cooler temperatures \citep{Levy1998, Attwater2010, Kua2011, Akoopie2016}. However, for hot origin-of-life scenarios, hydrothermal systems may be compelling venues for cyanosulfidic reaction networks like that of \citet{Patel2015}, reinforcing the utility of volcanism for prebiotic chemistry. %%%Remove?

\section{Conclusion \& Next Steps\label{sec:conc}}
Constraining the abundances of trace chemical species on early Earth is important to understanding whether postulated prebiotic pathways which are dependent on them could have proceeded. Here, we show that prebiotically-relevant levels of certain sulfidic anions are globally available in shallow, well-mixed aqueous reservoirs due to dissolution of sulfur-bearing gases that are volcanically injected into the atmosphere of early Earth. In particular, anions derived from SO$_2$ are available at $\geq 1\mu$M levels in non-acidic reservoirs for SO$_2$ outgassing rates corresponding to the modern Earth and higher. During episodes of intense volcanism, like the emplacement of basaltic fields like the Deccan Traps, SO$_2$-derived anions may be available at $\geq 1m$M levels for reservoirs buffered to pH$\geq 7$ (e.g., the modern ocean at pH$=8.2$) and at a temperature of $T=25^\circ$C, though sulfite disproportionation may have ultimately limited concentrations to the $\sim10 \mu$M level; better constraints on sulfite disproportionation reaction rates are required to constrain this possibility. At cooler temperatures, even higher concentrations of these anions would have been available. Formation of mineral precipitate should not inhibit sulfite concentrations until $\geq 1 m$M concentrations so long as the reservoir is not acidic, but might suppress sulfite levels in acidic waters. On the other hand, anions derived from H$_2$S would not have been available at micromolar levels across the plausible range of volcanic outgassing due to low solubility of H$_2$S and an unfavorable dissociation constant, and prebiotic chemistry invoking such anions must invoke local, specialized sources. Radiative transfer calculations suggests that NUV radiation will remain abundant at the planet surface for  $\phi_S\leq1\times10^{11}$ cm$^{-2}$ s$^{-1}$, but will be suppressed for  $\phi_S\geq3\times10^{11}$ cm$^{-2}$ s$^{-1}$; such epochs may be especially clement for surficial, UV-independent prebiotic chemistry. We applied our results to the case study of the proposed prebiotic reaction network of \citet{Patel2015}. The prebiotic plausibility of this network can be improved if it can be adapted to use SO$_2$-derived anions like HSO$_3^-$ or SO$_3^{2-}$ instead of HS$^-$, since the atmosphere is capable of supplying prebiotically-relevant levels of the former directly but more localized sources must be invoked for adequate supply of the latter. Coupled with the potential for volcanogenic synthesis of feedstock molecules like HCN and phospate \citep{Zahnle1986, Yamagata1991}, it appears that episodes of moderately intense volcanism ($\phi_S\approx1\times10^{11}$ cm$^{-2}$ s$^{-1}$) might have been especially clement for cyanosulfidic prebiotic chemistry which exploits SO$_2$-derived anions (e.g., HSO$_3^-$). Avenues for future work include simulating these scenarios experimentally and/or with a large general purpose aqueous geochemistry code, improving measurements of the sulfite disproportionation reaction rate constant, and further photochemical modelling to improve constraints on the expected concentrations of SO$_2$ and H$_2$S on early Earth.

\section{Acknowledgements}
We thank A. Levi for his insight regarding equilibrium chemistry, and D. Catling, I. Halevy, M. Claire, J. Szostak, I. Halevy, T. Bosak, and T. Vick for answers to many questions and/or comments on the manuscript. We especially thank J. Toner for extensive discussion and feedback. We thank M. Claire for sharing sulfur aerosol cross-sections for validation purposes. This research has made use of NASA's Astrophysics Data System Bibliographic Services, and the MPI-Mainz UV-VIS Spectral Atlas of Gaseous Molecules. S. R., Z. R. T., D. D. S., and J. D. S. gratefully acknowledge support from the Simons Foundation (S. R., Z. R. T., D. D. S.: grant no. 290360; S.R., grant no. 495062; J. D. S.: grant no. 290362).

\section{Author Disclosure Statement}
The authors declare no competing financial interests.

\clearpage
\bibliography{sulfidic_plausibility_v18.bib}   % name your BibTeX data base
\clearpage

\appendix
\section{Atmospheric Sulfur Speciation\label{sec:hutbl}}

We use the work of \citet{Hu2013} to connect the sulfur emission flux $\phi_S$ to the speciation of atmospheric sulfur. Table~\ref{tbl:compare_sulfur_hu} presents H$_2$S and SO$_2$ mixing ratios as a function of $\phi_S$ from \citet{Hu2013} (their Fig. 5, CO$_2$-dominated atmosphere case).

\begin{table}[h]
%\centering
\caption{Column-integrated mixing ratios of H$_2$S and SO$_2$ as a function of $\phi_S$ from \citet{Hu2013} (their Fig. 5, CO$_2$-dominated case). \label{tbl:compare_sulfur_hu} }
\begin{tabular}{p{3 cm}p{3 cm}p{3 cm}}
 \hline \noalign{\smallskip}
 $\phi_S$ (cm$^{-2}$~s$^{-1}$)& $r_{H_{2}S}$ & $r_{SO_{2}}$ \\
 \noalign{\smallskip}\hline\noalign{\smallskip}

$3\times10^{9}$ & $4\times10^{-10}$& $3\times10^{-10}$\\
$1\times10^{10}$ & $1\times10^{-9}$ & $9\times10^{-10}$ \\
$3\times10^{10}$ & $9\times10^{-9}$ & $3\times10^{-9}$ \\
$1\times10^{11}$ & $5\times10^{-8}$ & $7\times10^{-9}$ \\
$3\times10^{11}$ & $2\times10^{-7}$ & $1\times10^{-8}$ \\
$1\times10^{12}$ & $7\times10^{-7}$ & $3\times10^{-8}$ \\
$3\times10^{12}$ & $2\times10^{-6}$ & $8\times10^{-8}$ \\
$1\times10^{13}$ & $9\times10^{-6}$ & $3\times10^{-7}$ \\
\noalign{\smallskip}\hline
\end{tabular}
\end{table}

\section{Activity Coefficient Calculation\label{sec:appendix_activitycoeffs}}
This appendix describes the calculation of the activity coefficients of the ions involved in equilibria reactions for SO$_2$ and H$_2$S. 

We use the Extended Debye-Huckel (EDH) Theory to calculate activity coefficients ($\gamma_{i}$) for the ions in our study. EDH is valid for ionic strengths up to 0.1M, which is the highest ionic strength we consider, motivated by the fact that lipid vesicle formation is inhibited at ionic strengths above 0.1M \citep{Maurer2016}. 

Extended Debye Huckel theory states that:
\begin{align}
log{\gamma_i} = -Az_i^2 \frac{I^{0.5}}{1+B\alpha_i I^{0.5}}
\end{align}

where $A$ and $B$ depend on the temperature, density, and dielectric constant of the solvent (in our case water), and $\alpha_i$ is an ion-specific parameter. We took $A=0.5085$ M$^{-1/2}$ and $B=0.3281$ M$^{-1/2} \AA^{-1}$, corresponding to $T=25^\circ$C; Appendix~\ref{sec:appendix_temp} describes the sensitivity of our analysis to this assumption. Table ~\ref{tbl:activity} summarizes the $\alpha_i$ used in our study, taken from \citet{Misra2012}.  We were unable to locate a value of $\alpha_C$ for HS$_{2}$O$_{5}^{-}$, and consequently take $\gamma_{HS_{2}O_{5}^{-}}=1$ throughout (i.e., we do not correct for its activity). Since in our analysis the supply of SO$_2$ is not limited (the atmosphere is treated as an infinite reservoir), pKa$_{SO_{2}, 3}$ affects only the abundance of H$_2$SO$_5^-$, which is a trace compound in our analysis (see Fig.~\ref{fig:so2}).

\begin{table}[h]
%\centering
\caption{Values for the ion-specific parameter $\alpha$ (related to the hydration sphere of the ion) used to calculate activity coefficients. \label{tbl:activity}}
\begin{tabular}{p{2 cm}p{2cm}}
\hline\noalign{\smallskip}
Ion & $\alpha_i (\AA)$ \\
\noalign{\smallskip}\hline\noalign{\smallskip}
HSO$_3^-$ & 4.0 \\
SO$_3^{2-}$ & 4.5 \\
HS$^-$ & 3.5 \\
S$^{2-}$ & 5.0 \\
OH$^-$ & 3.5 \\
H$^+$ & 9.0 \\
\noalign{\smallskip}\hline
\end{tabular}
\end{table}

 Table ~\ref{tbl:activity_coeff} shows the activity coefficients for the relevant ions at the two ionic strengths considered in our study:

\begin{table}[h]
%\centering
\caption{Per-ion activity coefficients for different ionic strengths.\label{tbl:activity_coeff}}
\begin{tabular}{p{1.7 cm}p{1.7 cm}p{1.7 cm}}
 \hline\noalign{\smallskip}
Ion & I=0 M & I=0.1 M  \\
\noalign{\smallskip}\hline\noalign{\smallskip}
HSO$_3^-$  & 1.0 & 0.770 \\
SO$_3^{2-}$ &  1.0 &  0.364 \\
HS$^-$ & 1.0 & 0.762 \\
S$^{2-}$& 1.0 & 0.377 \\
OH$^{-}$ & 1.0 & 0.762 \\
H$^+$ & 1.0 & 0.826 \\
\noalign{\smallskip}\hline
\end{tabular}
\end{table}

\section{Sensitivity of Henry's Law Constants to Salinity\label{sec:appendix_henry}}
This appendix describes our assessment of the sensitivity of the Henry's Law coefficients for SO$_2$ and H$_2$S to salinity. 

We account for the effect of salinity on $H_G$ using the Schumpe-Sechenov method, as outlined in \citet{Burkholder2015}: 

\begin{align}
\log{H_0/H}=\Sigma_i (h_i+h_G)*c_i,
\end{align}

where $H_0$ is the Henry's Law constant in pure water, H is the Henry's law constant in saline solution, $c_i$ is the concentration of the ion $i$, $h_i$ is an ion-specific constant, and $h_G$ is a gas-specific constant. $h_G$ is temperature dependent, via $h_G=h_{0}+h_T (T-298.15 K)$. NaCl is the dominant salt in Earth's oceans; we approximate NaCl as the sole source of salinity in our calculations. Table~\ref{tbl:henry} summarizes the values of these parameters used for this study, all taken from the compendium of \citet{Burkholder2015}. We were unable to locate a value for $h_T$ for H$_2$S in our literature search, and assumed $h_T=0$ for this case.

\begin{table}[h]
%\centering
\caption{Parameters used to estimate the dependence of Henry's Law constants on [NaCl]. \label{tbl:henry}}
\begin{tabular}{p{2.5 cm}p{1.5 cm}p{1.5 cm}p{1.5 cm}p{1.5 cm}}
 \hline\noalign{\smallskip}
 Parameter & H$_2$S & SO$_2$ &Na$^+$ & Cl$^-$\\
 \noalign{\smallskip}\hline\noalign{\smallskip}
$H_0$ (M/bar) & 0.101 & 1.34 & -- & -- \\
$h_{0}$ (M$^{-1}$) & -0.0333 & -0.0607 & -- & -- \\
$h_T$ (M$^{-1}$) & 0$^a$& 0.000275 & -- & --\\
$h_i$  (M$^{-1}$)& -- & -- & 0.1143 & 0.0318\\
 \noalign{\smallskip}\hline
\end{tabular}
\end{table}

\begin{figure}[h]
\centering
\includegraphics[width=.8\linewidth]{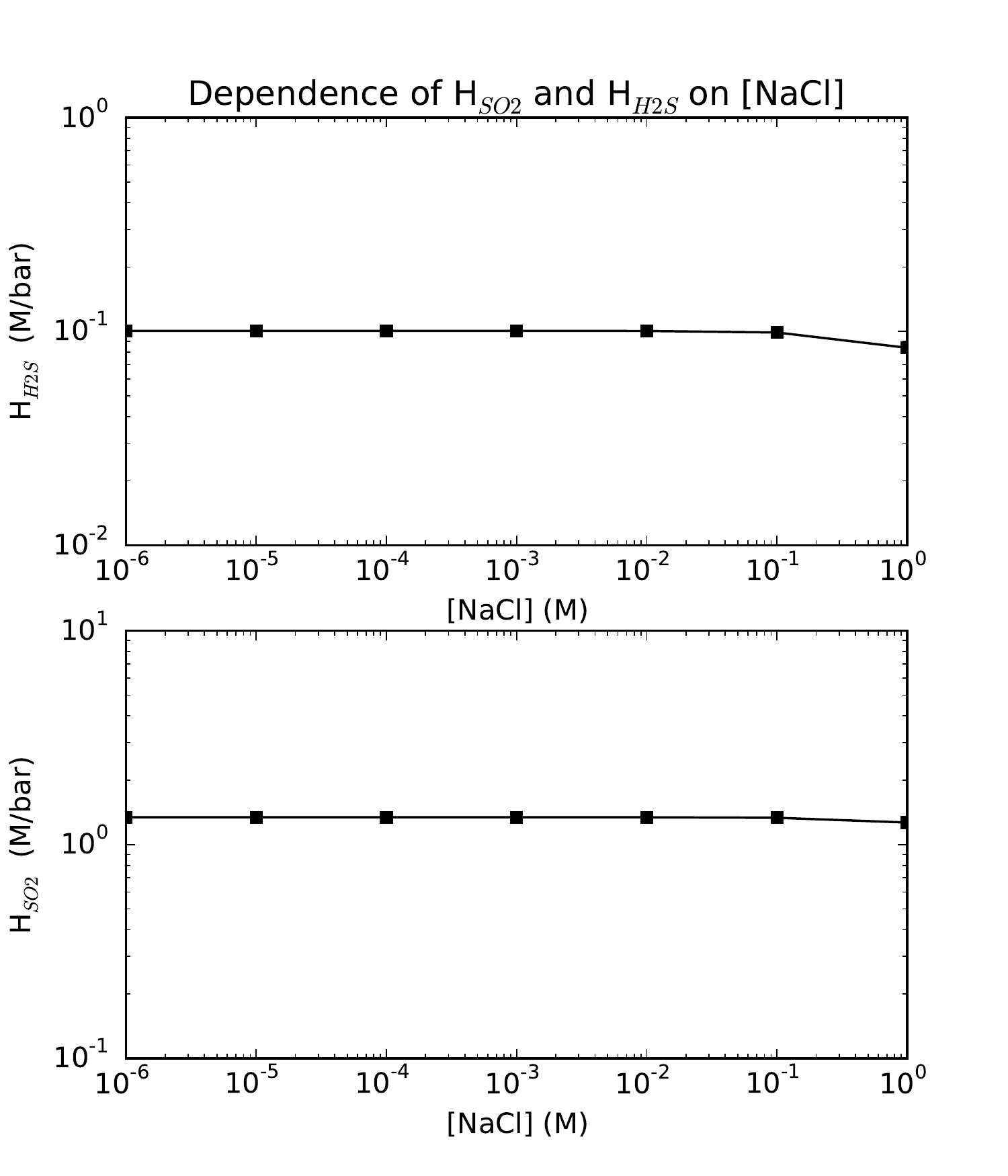}
\caption{Dependence of Henry's Law constants for H$_2$S and SO$_2$ on [NaCl], calculated using the formalism from \citet{Burkholder2015}. $H_{H_{2}S}$ and $H_{SO_{2}}$ are insensitive to [NaCl] for [NaCl]$<1$M.\label{fig:henry_nacl_dep}}
\end{figure}

The Henry's Law constants for these gases as a function of [NaCl] at $T=298.15$K is show in Fig.~\ref{fig:henry_nacl_dep}. In this study, we consider ionic strengths $I\leq0.1$M, corresponding to [NaCl]$\leq0.1$M. At such levels, salinity has a negligible effect on Henry's Law solubility, and we consequently neglect it in our calculations.

\section{Sensitivity of Analysis to Temperature\label{sec:appendix_temp}}
This appendix describes our assessment of the sensitivity of our calculations to the temperature of the aqueous reservoir in which the equilbrium chemistry proceeds.

\subsection{Sensitivity of Henry's Law Constants to Temperature\label{sec:appendix_henry_temp}}
\begin{figure}[h]
\centering
\includegraphics[width=.8\linewidth]{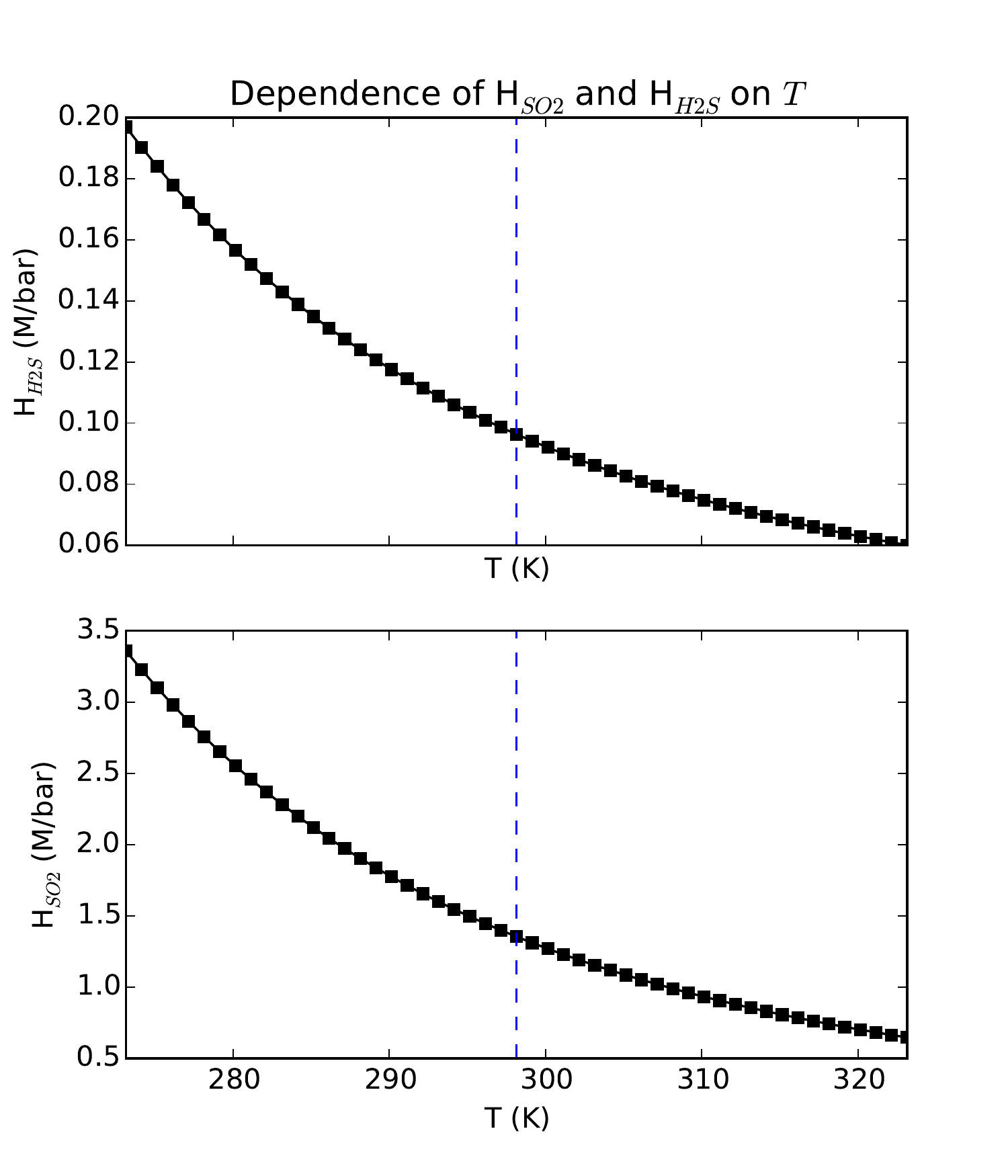}
\caption{Temperature dependence of Henry's Law constants for H$_2$S and SO$_2$, calculated using the formalism from \citet{Burkholder2015}. Varying the temperature by 25K relative to the reference temperature of 298.15K (blue line) affects the value of $H_{H_{2}S}$ and $H_{SO_{2}}$ by less than a factor of 2.5. \label{fig:henry_temp_dep}}
\end{figure}

We calculated the effect of temperature on Henry's Law using the three-term empirical fit outlined in \citet{Burkholder2015}, i.e. $\ln(H)=A+B/T+C\ln(T)$, where $H$ is in units of M/atm and $A$, $B$, and $C$ are gas-specific coefficients of an empirical fit. The values of these coefficients for H$_2$S and SO$_2$ were taken from \citet{Burkholder2015} and are summarized in Table~\ref{tbl:henry_temp}. $H(T)$ for H$_2$S and SO$_2$ is plotted in Fig.~\ref{fig:henry_temp_dep}. For temperatures ranging from $0-50^\circ$ ($273.15-323.15$ K), the Henry's Law constants vary by less than a factor of 2.5 relative to their values at $25^\circ$C (293.15K), which is small compared to the order-of-magnitude variations in concentration we focus on in this study. We also estimated the temperature dependence using the van't Hoff equation as outlined in \citet{Sander2015}, and obtained similar results.

\begin{table}[h]
%\centering
\caption{Parameters used to estimate dependence of Henry's Law constant on temperature\label{tbl:henry_temp}}
\begin{tabular}{p{2 cm}p{1.5 cm}p{1.5 cm}}
 \hline\noalign{\smallskip}
 Parameter & H$_2$S & SO$_2$ \\
\noalign{\smallskip}\hline\noalign{\smallskip}
$A$ & -145.2 & -39.72 \\
$B$& 8120 & 4250\\
$C$ & 20.296 & 4.525\\
 \noalign{\smallskip}\hline
\end{tabular}
\end{table}

\subsection{Sensitivity of Reaction Rates to Temperature\label{sec:appendix_ratetemp}}
In order to assess the temperature dependence of acid dissociation pKa's, we use the Van't Hoff Equation: 
\begin{align}
\frac{\partial[lnK_0]}{\partial T} = \frac{\Delta H_0}{RT^{2}}
\end{align}

Where $K_0$ is the equilibrium constant, $T$ is temperature, $\Delta H_0$ is the change in enthalpy, and $R=8.314 \times 10^{-3} kJ/mol/K$. Solving this differential equation, assuming temperature-invariant enthalpy of solution\footnote{We expect this assumption to be reasonable because of the limited range of temperatures that are plausible for our surface aqueous reservoir scenario.}, gives:
\begin{align}
ln(K_2) = ln(K_1) + \frac{- \Delta H_0}{R} \left( \frac{1}{T_2} - \frac{1}{T_1} \right)  
\end{align}

With this equation, the acid dissociation constant $K_2$ can be estimated at a given temperature $T_2$, provided its value $K_1$ is known at a reference temperature $T_1$. $\Delta H$ is the change in enthalpy of the reaction, given by:

\begin{align}
\Delta H = \sum \Delta H_f^\circ (products) - \sum \Delta H_f^\circ (reactants)
\end{align}

The enthalpies of formation for the products and reactants of the first two acid dissociation reactions for H$_2$S and SO$_2$ are taken from \citet{CRC90}, and are shown in Table~\ref{tbl:enthalpy}. Note that $\Delta H_f^\circ=0$ for H$^+$, by definition. We were unable to locate an enthalpy of formation for H$_2$SO$_5^-$, and consequently are unable to calculate the temperature-dependence of pKa$_{SO_{2}, 3}$. Since in our analysis the supply of SO$_2$ is not limited (the atmosphere is treated as an infinite reservoir), pKa$_{SO_{2}, 3}$ affects only the abundance of H$_2$SO$_5^-$, which is a trace compound in our analysis (see Fig.~\ref{fig:so2}). 

\begin{table}[h]
%\centering
\caption{Enthalpies of formation used in the Van't Hoff Equation calculation. \label{tbl:enthalpy}}
\begin{tabular}{p{1.5 cm}p{3 cm}}
\hline\noalign{\smallskip}
 Molecule& $\Delta H_f^\circ$ (kJ/mol)  \\
\noalign{\smallskip}\hline\noalign{\smallskip}
HSO$_3^-$  &  -626.2  \\
SO$_3^{2-}$ &  -635.5 \\
SO$_2$ & -296.8 \\
HS$^-$ & -17.6 \\
S$^{2-}$& 33.1 \\
H$_2$S & -20.6 \\
H$_2$O &  -285.8 \\
\noalign{\smallskip}\hline
\end{tabular}
\end{table}

Using these values and the Van't Hoff equation, we calculated the temperature dependence of the first two acid dissociation constants for aqueous H$_2$S and SO$_2$. Table ~\ref{tbl:temp_dep} shows these pKa's for SO$_2$ and H$_2$S at 0, 25, and 50$^\circ$C. 

\begin{table}[h]
%\centering
\caption{$pKa$ at temperatures of 0$^\circ$C, $25^\circ$C, and 50$^\circ$C and reaction enthalpies\label{tbl:temp_dep}}
\begin{tabular}{p{3 cm}p{3 cm}p{3 cm}p{3cm}p{3cm}}
\hline\noalign{\smallskip}
 & T=0 $^\circ$ C  &  T=25 $^\circ$ C & T=50 $^\circ$ C  & $\Delta H^\circ_{rxn}$ (kJ/mol) \\
 \noalign{\smallskip}\hline\noalign{\smallskip}
H$_2$S, $pKa_1$ &  7.098 & 7.05 & 7.009 & 3.0 \\
H$_2$S, $pKa_2$ &  19.81 & 19.0 & 18.31   & 50.7 \\
SO$_2$,  $pKa_1$ &  1.160 & 1.86 & 2.452 & -43.6 \\
SO$_2$,  $pKa_2$ &  7.051 & 7.2 & 7.326 & -9.3 \\
\noalign{\smallskip}\hline
\end{tabular}
\end{table}

The variation in pKa is negligible for all reactions except the first dissociation of SO$_2$; pKa$_{SO_{2},1}$ increases significantly with temperature. This implies that in non-acidic solutions, the concentrations of SO$_2$-derived anions should decrease, and conversely that as temperature decreases they should increase.

\subsection{Sensitivity of Activity Coefficients to Temperature\label{sec:appendix_actcoeff_temp}}
Temperature dependence enters the calculation of the activity coefficients through the parameters $A$ and $B$ (see Appendix~\ref{sec:appendix_activitycoeffs} for details). For water, at $T=0^\circ$C, $A=0.4883$ M$^{-1/2}$ and $B=0.3241$ M$^{-1/2} \AA^{-1}$; at $T=25^\circ$C, $A=0.5085$ M$^{-1/2}$ and $B=0.3281$ M$^{-1/2} \AA^{-1}$; and at $T=50^\circ$C, $A=0.5319$ M$^{-1/2}$ and $B=0.3321$ M$^{-1/2} \AA^{-1}$ \citep{Misra2012}. From $T=0-50^\circ$C, the activity coefficients varied by $<8\%$ for $I\leq 0.1$.

\subsection{Overall Sensitivity of Analysis to Temperature\label{sec:appendix_temp_combined}}
We evaluated the overall sensitivity of our analysis to our assumption of $T=25^\circ$C by repeating our analysis at $T=0^\circ$ and $T=50^\circ$, and including the effects of temperature on Henry's Law constant, the reaction pKa's, and the activity coefficients, simultaneously. Across the tested range, temperature had a negligible impact on the abundances of the H$_2$S-derived anions, but a significant impact on the abundances of the SO$_2$-derived anions. $H_{SO_{2}}$ decreases with temperature, and pKa$_{SO_{2},1}$ increases with temperature; both effects serve to increase the concentration of HSO$_3^-$ and its derivatives in non-acidic (pH$>2.5$) waters. At $T=0^\circ$C, HSO$_3^-$ and SO$_3^{2-}$ concentrations are an order of magnitude higher than at $T=25^\circ$C. Similarly, at $T=50^\circ$C, HSO$_3^-$ and SO$_3^{2-}$ concentrations are an order of magnitude lower than at $T=25^\circ$C. Our overall conclusions are robust to these variations. However, this study does imply that   significantly higher concentrations of SO$_2$-derived anions are available to prebiotic chemistry in cooler waters, and inversely that hotter waters would have access to lower levels of SO$_2$-derived anions.

\end{document}